\documentclass{aa}

\input{psfig.tex}
\def\gsim{\;\lower.6ex\hbox{$\sim$}\kern-7.75pt\raise.65ex\hbox{$>$}\;}
\def\lsim{\;\lower.6ex\hbox{$\sim$}\kern-7.75pt\raise.65ex\hbox{$<$}\;}
\begin{document}

%

%
\title{ Abundance Analysis of Turn-Off and early Subgiant Stars 
in the Globular
Cluster 47 Tuc (NGC 104)\thanks{Based on data collected at the European
Southern Observatory, Chile, in the course
of the ESO-Large program 165.L-0263} }

\author{
E. Carretta\inst{1},
R.G. Gratton\inst{1},
A. Bragaglia\inst{2},
P. Bonifacio\inst{3}
\and
L. Pasquini\inst{4}
}

\authorrunning{E. Carretta et al.}
\titlerunning{Abundance analysis in subgiants and dwarfs of 47 Tuc}


\offprints{E. Carretta, carretta@pd.astro.it}

\institute{
INAF - Osservatorio Astronomico di Padova, Vicolo dell'Osservatorio 5, I-35122
 Padova, Italy
\and
INAF - Osservatorio Astronomico di Bologna, Via Ranzani 1, I-40127
 Bologna, Italy
\and
INAF - Osservatorio Astronomico di Trieste, via Tiepolo 11, I-34131 Trieste, Italy
\and
ESO - European Southern Observatory
  }

\date{Received 22/9/2003; accepted 5/11/2003}

\abstract{We used the UVES spectrograph on Kueyen (VLT UT2) to perform the
abundance analysis of stars at the base of the giant branch (RGB) and at the
main sequence turn-off in the globular cluster 47 Tuc (NGC 104). High
dispersion spectra ($R\gsim 40,000$) for 3 dwarfs and 9 subgiants were
analyzed. We found an overall [Fe/H] value of $-0.67\pm0.01\pm0.04$,
$rms$ = 0.05 dex. The mean value obtained from stars at the base of the RGB is
virtually the same as that obtained from the dwarfs. Na and O abundances show a
star-to-star scatter and are anti-correlated, not unlike 
previous finding
in more metal-poor clusters. The extension of the anticorrelation is not
as extreme as found in other clusters, but it is clearly present in unevolved
stars. Al abundances do not show a significant spread; this could be indicative 
that p-captures in high temperature regions are less likely to occur in
metal-rich clusters. The
[$\alpha$/Fe] ratios suggest 
a slight excess of Ti with respect to field
stars of similar metallicities. 
Finally, we found an enhanced odd-even
effect for Fe-group elements. In particular, 
the [Mn/Fe] ratio is almost 0.2 dex
deficient with respect to the mean value of the dissipative component of field
stars in the solar neighborhood. While the theoretical and observational
framework for Mn is still poorly understood, the chemistry of 47 Tuc, coupled
with its kinematics and its somewhat younger age, might give hints favouring 
an origin in a formely independent, possibly larger, subsystem of our Galaxy.
}


\maketitle

\section{INTRODUCTION}

The galactic globular 
clusters (GCs) are relevant for a large set of astrophysical
problems, ranging from stellar evolution to cosmology.
Their chemical compositon is a fundamental probe of the
early chemical evolution in the galactic environment. However, before using
globular clusters as tracers of the pristine chemical conditions in the
Galaxy, one has to disentangle the original compositions 
from any effect which
could have resulted in a pollution of their photospheres.
Possible effects include (i) possible self-enrichment of the protoclouds which
gave origin to the GCs (Cayrel 1986, Brown et al. 1991, 1995); (ii)  changes in
abundances derived from the internal evolution of the stars (both canonical or
non-canonical mixing, see Kraft 1994, Gratton et al. 2000); and (iii)
modifications of the composition due to a possible second generation of stars
born
from the low velocity wind of intermediate mass globular cluster stars
(Parmentier et al. 1999; Parmentier and Gilmore 2001).

A growing body of evidences shows that the usual assumption
of globular
cluster stars
being exactly coeval and chemically homogeneous (the basis of their
classification as a pure simple stellar population) is breaking down.
Useful records of the first phases of formation and early
evolution of GCs are hidden into the photospheric abundances of stars presently
observed. Once correctly interpreted, they may provide an understanding  of
the origin and evolution of GCs which may not be achieved
by  other  means.

The most noticeable of these
anomalies are those involving the light element C, N, O (for exhaustive
reviews see e.g. Smith 1987; Smith \& Mateo 1990; Kraft 1994 and
references therein), as detected from 
molecular abundances of CN, CH, NH and
OH, 
often accompanied by correlated 
anomalies in atomic species such as O, Na, Al and
Mg (see the survey by the Lick-Texas group).

The observed abundance pattern may be explained by H-burning at high
temperatures, where reactions and products of the CNO, NeNa and MgAl
cycles occur. Denisenkov \&
Denisenkova (1990), Langer et al. (1993) have shown that production  of
$^{23}$Na and $^{27}$Al from proton captures on $^{22}$Ne and $^{25}$Mg is
possible even in low mass stars, in regions close to the ON
burning shell.

To explain the complex observational picture, 
there are presently two main lines of thought.
The first (see Cottrell \& Da Costa 1981; Brown \& Wallerstein 1992),
considers primordial inhomogeneities, i.e. already existing in the gas from
which the cluster stars formed, or released from the very early generation of
massive stars born and evolved in GCs. This scenario is based on the belief
that Na and Al cannot be synthetized in the low mass stars 
presently observed
in GCs, so that the overabundances of these elements, and the correlated
CN-enhancements, 
must be due to the yields of more massive stars that polluted the
next generation of cluster stars.

The second scenario explains the CN and anticorrelated anomalies as due to
internal deep mixing, that brings up
to the surface fresh $^{14}$N from CN process,
depleting at the same time the abundance of $^{12}$C of the surface layers. 
The constant decrease
of the isotopic ratio $^{12}$C/$^{13}$C up to 
values near the equilibrium ratio
of 3.5 (see the review of Suntzeff 1993) supports this mechanism.
All stars experience a mixing phase (the standard first dredge-up) during the
first ascent on the giant branch; however, to explain the fact that only some
of the stars are CN-strong, one has to assume the existence of a mechanism of
extra mixing.

Following the model of Sweigart \& Mengel (1979), this additional mechanism
could be represented by 
meridional circulation currents driven by core rotation. In order to explain 
the observations, these currents should be able
to reach the region of the ON process, bringing up to the surface processed
N-rich and O-poor material that contributes to the CN enhancements even in the
case of large C depletions.
However, in the model of Sweigart \& Mengel (1979) a gradient of mean molecular
weight $does$ inhibite the formation of circulation currents  (Tassoul \& Tassoul
1984), and so the additional mixing, below a  critical level (log L/L$_\odot$
$\sim$ 2.2) on the RGB. This is the point of maximum penetration of the
convective envelope, succesively  crossed by the H-burning shell propagating
outward (the so-called RGB-bump).

The observational evidence
shows the opposite:
anomalous CN abundances exist 
at considerably lower luminosities than predicted.
Furthermore, a radial
dependence in the distributions of CN-strong and CN-weak stars (the first being
more centrally concentrated) seems to be present in 47 Tuc and possibly in
other clusters (Norris 1987), and this is not easily explained by the
mixing scenario.

Recently  we definitively settled this issue by
finding a clear Na-O anticorrelation among unevolved (turn-off) stars in the
globular cluster NGC 6752
(Gratton et al. 2001, hereafter Paper I). These stars do not meet the
constraints of high central temperature and extended envelope required to
self-produce and bring to the surface the synthetized proton-capture elements.
The same seems to hold also in the more metal-poor cluster NGC 6397 (Carretta
et al. 2003a, in preparation).

The Na-O anticorrelation is apparently limited to the dense environments of
GCs: there are almost no stars showing  CN-strong or Na-rich composition in the
field, where the chances for pollution are much less than in the denser GCs.
In fact, Gratton et al. (2000) convincingly showed that the abundances of
elements heavier than CN (e.g. Na and O) are unaltered in field stars of
whatever evolutionary stages, from pre-main sequence up to the horizontal
branch (HB) phase.

There are several reasons which make
47 Tuc a preferred target for studying the
complex issue of chemical anomalies in globular cluster stars:
\begin{itemize}
\item among the few most nearby clusters, it presents a reasonably low value of
the interstellar extinction

\item 47 Tuc is traditionally adopted as the classical template of the
high-metallicity tail of the metal distribution of globular clusters

\item as a typical metal-rich cluster, 47 Tuc is considered a pillar of
the globular cluster distance scale based on the main sequence fitting method
(see e.g. Carretta et al. 2000, Gratton et al. 2003a)

\item this cluster is very well studied, by the photometric point of view, and
large databases in different photometric systems exist. However, despite its
proximity it has not been extensively studied with high-resolution
spectroscopy. Instead, a sparse number of its giant stars
have been observed, mostly for calibrating purposes, as in the Norris and
Da Costa (1995) study of $\omega$ Cen

\item finally, the relevance of studying 47 Tuc is evident, since most
mixing mechanisms are less and less efficient as the global metal abundance
increases, due to the strengthening of the molecular weight barrier  (Sweigart
and Mengel 1979). Hence, in metal-rich environments it should be easier to
investigate the  primordial component of the so-called chemical anomalies in
globular cluster stars.
\end{itemize}

Several studies at medium and low resolution have
been performed in the past. They were aimed to
extract informations from CN and CH band strengths 
in main sequence and turn-off
stars of 47 Tuc (Hesser 1978; Hesser and Bell
1980; Bell, Hesser and Cannon 1983, Briley, Hesser and Bell 1991; Briley et al.
1996; Cannon et al. 1998 and references therein). 
Pushing to the limit the size and
limit magnitude of the samples, 
these investigations uncovered many features of
the CH and CN distributions also among unevolved dwarfs in this cluster.

The favourable location of the ESO VLT complex in the southern emisphere,
coupled with the high efficiency of the UVES spectrograph, now allow us to make
a significant step forward and obtain high-res spectra at a S/N adequate for a
precise, detailed abundance analysis of several elements in 47 Tuc dwarfs.

\begin{figure}
\psfig{figure=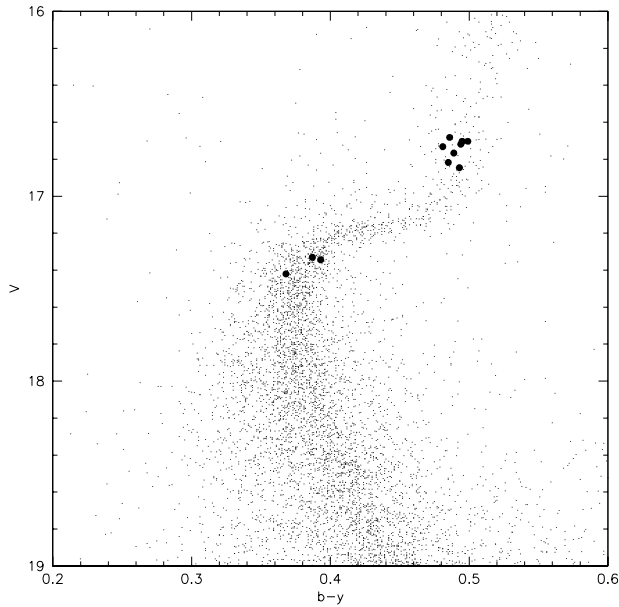,width=8.8cm,clip=}
\caption[]{Positions of program stars (large symbols) on the Str\"omgren
colour-magnitude diagram of 47 Tuc. Photometry
is from Grundahl et al. (1999).}
\label{f:selstar}
\end{figure}

\section{OBSERVATIONS}

Observations were carried out in three different runs (September 2000; August
and October 2001) with the spectrograph UVES at VLT-UT2 of the Paranal ESO
Observatory; each run consisted in 6 nights assigned to the ESO Large
Program 165.L-0263 and its following time allocations.
Due to poor weather conditions in the September 2000 run, we were not able to
complete the acquisition of the spectra in 47 Tuc. Bad weather/seeing plagued
also the August 2001 and partially the October 2001 runs. In total, 5 candidate
turn-off stars and 9 candidate subgiant stars were observed.
However, one of the turn-off stars turned out to be not-member, both from radial
velocity (see Lucatello and Gratton 2003) and elemental abundances, and
for another we were able to gather only a low-S/N ($\sim 15$) spectrum, useless
apart from the membership.
Hence we have useful observational material (high-resolution, high S/N spectra)
only for 9 stars at the base of the RGB and 3 stars at the main
sequence turn-off in 47 Tuc.

Following Paper I, we tried to observe stars likely having both
strong and weak CN bands, selecting objects on the basis of 
Str\"omgren photometry, using the index $c_1$ (increasing with the strength
of CN bands, Grundahl et al. 1999).

Selected stars are indicated in Figure~\ref{f:selstar}, and their relevant data
are listed in Table~\ref{t:reldata}.

\begin{table*}
\caption[]{Data for observed stars in 47 Tuc (identifications from 
Grundahl et al. 1999);}
\begin{tabular}{rcclcccccccc}
\hline
Star   &$S/N$& $v_r$       &observing  & B & V & u & b & v& y & RA(2000)& DEC(2000)\\
       &     & km s$^{-1}$ &run        &   &   &   &   &  &  & &\\
\hline
\multicolumn{12}{c}{Subgiants}      \\
   435 & 40 & $-$20.2 $\pm 0.6 $  & Oct 2001 & 17.540 &16.739 &19.022 & 17.213 & 17.960 & 16.732 & 00:21:13.65&  -72:03:32.41\\
   456 & 40 & $-$16.1 $\pm 0.3 $  & Oct 2001 & 17.537 &16.746 &19.016 & 17.256 & 17.989 & 16.767 & 00:21:30.42&  -72:05:35.95\\
   433 & 40 & $-$23.0 $\pm 0.2 $  & Oct 2001 & 17.502 &16.723 &18.993 & 17.198 & 17.945 & 16.703 & 00:21:49.24&  -72:02:54.06\\
   478 & 30 & $-$20.4 $\pm 0.3 $  & Oct 2001 & 17.609 &16.827 &19.131 & 17.339 & 18.076 & 16.846 & 00:21:08.17&  -71:58:47.54\\
201600 & 40 & $-$9.5  $\pm 0.4 $  & Oct 2001 & 17.470 &16.669 &19.001 & 17.168 & 17.920 & 16.682 & 00:21:52.55&  -72:05:27.66\\
   429 & 35 & $-$19.3 $\pm 0.2 $ & Sept 2001 & 17.499 &16.699 &19.022 & 17.202 & 17.953 & 16.703 & 00:20:18.60&  -72:01:11.40\\
201075 & 40 & $-$10.2 $\pm 0.2 $ & Sept 2001 & 17.609 &16.817 &18.987 & 17.302 & 18.013 & 16.817 & 00:22:00.40&  -72:05:59.40\\
206415 & 40 & $-$21.8 $\pm 0.3 $ & Sept 2001 &        &       &18.940 & 17.213 & 17.932 & 16.719 & 00:19:36.40&  -71:58:57.00\\
   482 & 45 & $-$0.6  $\pm 0.3 $  & Aug 2001 & 17.648 &16.816 &19.156 & 17.352 & 18.097 & 16.842 & 00:21:10.88&  -72:04:15.16\\

\multicolumn{12}{c}{Dwarfs}         \\
1012   & 30 & $-$22.8 $\pm 0.2 $  & Oct 2001 & 17.942 &17.361 &19.053 & 17.737 & 18.251 & 17.344 & 00:21:26.27&  -72:00:38.73\\
1081   & 45 & $-$16.1 $\pm 0.2 $  & Aug 2001 & 17.962 &17.375 &19.099 & 17.788 & 18.282 & 17.420 & 00:21:03.82&  -72:06:57.74\\
 975   & 40 & $-$20.2 $\pm 0.2 $ & Sept 2000 & 17.919 &17.322 &19.068 & 17.717 & 18.247 & 17.330 & 00:20:52.30&  -71:58:01.80\\
\hline
\end{tabular}
\begin{list}{}{}
\item[] Values of S/N are estimated at 6700~\AA; heliocentric radial 
velocities are from Lucatello \& Gratton (2003).
\end{list}                                     
\label{t:reldata}
\end{table*}

Like for NGC 6397 and NGC 6752 (see Paper I), data were acquired using the
dichroic beamsplitter \#2 at UVES. The CD2, centered at 420 nm (with
spectral coverage $\lambda\lambda$ 356-484 nm), was used in the blue arm of the
spectrograph; and the CD4 centered at 750 nm (covering $\lambda\lambda$
555-946 nm) was adopted in the red arm. Slit length was always 8 arcsec;
the slit width was mostly set at 1 arcsec (corresponding to a resolution of
43000 at the order centers). However, this value was slightly modified downward
or upward in a few cases, according to the seeing conditions.

Typical exposure times were 2 hours for each subgiant and about 4 hours for
each turn-off star, splitted in multiple exposures. Values of the S/N per pixel
(about 5 per resolution element) measured at about 6700~\AA\ are listed in
Table~\ref{t:reldata}.

\section{ATMOSPHERIC PARAMETERS AND IRON ABUNDANCES}

\subsection{Atmospheric parameters}

\begin{figure}
\psfig{figure=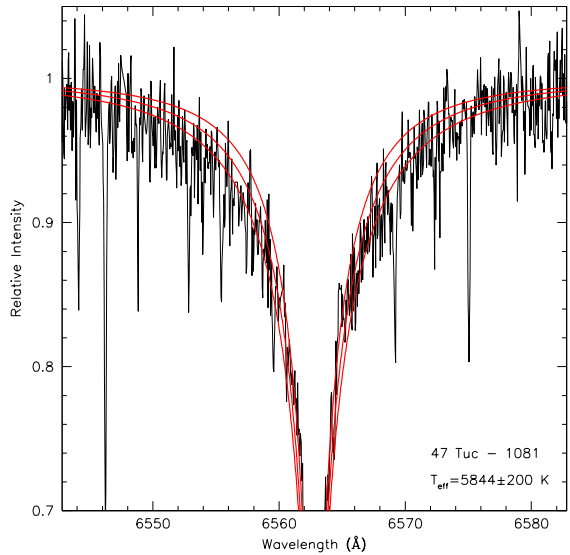,width=8.8cm,clip=}
\caption[]{Derivation of temperature from the H$\alpha$ profile for star
1081 in 47 Tuc. Thin line: observed profile; thick lines: expected
profiles computed for T$_{\rm eff}$ = 5644, 5844, and 6044 K. In the fitting we
used only the regions where the relative intensity is $> 0.9$.}
\label{f:fithalfa}
\end{figure}

Following the procedure adopted in Paper I, we compared effective temperatures
from observed colours (both Johnson $B-V$ and Str\"omgren $b-y$ were used) with
spectroscopic temperatures derived from comparison of Balmer line profiles (namely
H$\alpha$) with those obtained from spectral synthesis.
An example of the fit is shown in Figure~\ref{f:fithalfa} for star 1081.

Since (i) these last temperatures are reddening-free and (ii) they are derived
following a uniform procedure also for field stars (see Paper I), this
approach allows us to compare directly our derived abundances for 47 Tuc with
those obtained for field stars. This is a key requisite for accurate distances
from the Main Sequence Fitting Method (see Gratton et al. 2003a)\footnote{The
adopted temperature scale for field subdwarfs is very close to that of Alonso et
al.(1996)}.

However, some differences do exist with respect to the analysis performed for
stars in NGC 6397 and NGC 6752.

\begin{itemize}

\item we adopted individual T$_{\rm eff}$'s for subgiant stars, but
a unique value (the average of individual ones) for the 3 stars at
the main-sequence turn-off. This procedure was
devised as the one providing the best agreement with temperatures from line
excitation

\item we used new enhancement factors for collisional damping. These are
described in detail in Gratton et al. (2003b).

\item a new task was designed to iteratively clean the set of lines,
disregarding outliers that yield abundances
differing more than 2.5 $\sigma$ from the average abundance of the remaining
lines.

\end{itemize}

Values of the surface gravity were derived (as in Paper I) from the location of
stars in the colour magnitude diagram, when an age of 14 Gyr and corresponding
masses were assumed.

Estimates of the microturbulent velocity $v_t$ for each star were derived, as
usual, by eliminating trend of abundances with expected line strengths; to this
purpose, we used abundances of Fe I, for which a large number of lines were
measured in each star. At variance with Paper I, we found that, for stars at the
base of the giant branch, the star-to-star scatter in Fe abundances was reduced
if individual values of microturbulent velocity were used.

Finally, the overall model metallicity [A/H] was chosen as that of the model
atmospheres extracted from the grid of ATLAS models
with the overshooting option
switched off computed by F. Castelli
\footnote{available from 
{\tt http://kurucz.harvard.edu/}}, 
that best reproduce the measured equivalent widths (EW).

Final adopted atmospheric parameters are listed in Table~\ref{t:atmpar}.

\begin{table*}
\caption[]{Adopted atmospheric parameters and derived Iron abundances 
for observed stars in 47 Tuc }
\begin{tabular}{rccccrccrcc}
\hline
Star   &  T$_{\rm eff}$ & $\log$ $g$ & [A/H] & $v_t$  & n &[Fe/H]I & $rms$ & n & [Fe/H]II & $rms$\\
       &     (K)        &  (dex)     &  (dex)& (km s$^{-1}$) \\
\hline
\multicolumn{11}{c}{Subgiants}      \\
   435 &  5190  & 3.84 & $-$0.63 &  0.00 &55 & $-$0.64 & 0.17 & 6 & $-$0.48 & 0.17 \\
   456 &  5142  & 3.84 & $-$0.68 &  0.50 &51 & $-$0.71 & 0.11 & 6 & $-$0.59 & 0.09 \\
   433 &  5106  & 3.84 & $-$0.74 &  1.05 &55 & $-$0.78 & 0.15 & 6 & $-$0.57 & 0.11 \\
   478 &  5118  & 3.84 & $-$0.56 &  0.00 &59 & $-$0.59 & 0.20 & 6 & $-$0.65 & 0.15 \\
201600 &  5160  & 3.84 & $-$0.61 &  0.70 &52 & $-$0.65 & 0.12 & 9 & $-$0.63 & 0.13 \\
   429 &  5081  & 3.84 & $-$0.61 &  1.04 &67 & $-$0.65 & 0.16 &10 & $-$0.53 & 0.19 \\
201075 &  5165  & 3.84 & $-$0.64 &  0.70 &48 & $-$0.68 & 0.10 &10 & $-$0.59 & 0.14 \\
206415 &  5112  & 3.84 & $-$0.67 &  1.05 &58 & $-$0.70 & 0.13 & 8 & $-$0.61 & 0.15 \\
   482 &  5090  & 3.84 & $-$0.62 &  0.84 &62 & $-$0.62 & 0.09 & 9 & $-$0.36 & 0.14 \\

\multicolumn{11}{c}{Dwarfs}         \\
1012   &  5832  & 4.05 & $-$0.65 &  1.07 &38 & $-$0.64 & 0.21 & 7 & $-$0.66 & 0.09 \\
1081   &  5832  & 4.05 & $-$0.66 &  1.07 &41 & $-$0.64 & 0.19 & 9 & $-$0.74 & 0.11 \\
 975   &  5832  & 4.05 & $-$0.62 &  1.07 &39 & $-$0.64 & 0.18 & 6 & $-$0.67 & 0.19 \\
\hline
\end{tabular}
\label{t:atmpar}
\end{table*}

\subsection{Equivalent widths}

Automatic equivalent width  measurements of spectral lines were performed
exploiting a recent version of the spectrum analysis package developed in
Padova and partially described in Bragaglia et al. (2001) and Carretta et al.
(2002). 
The number of useful lines depends on the metallicity of star
and on the $S/N$. The metal abundance being the same for dwarfs and early
subgiants in 47 Tuc, we were able to measure similar numbers of Iron lines for
the two groups, i.e. from 50 to 60 Fe I lines for the SGB and from 30 to 40 Fe
I lines for dwarfs, and about 10 Fe II lines in each group, securing a good
database with statistical significance for the following analysis.

From the classical formula by Cayrel (1989) widely used to give an estimate  of
the error in measured $EW$ as a function of the full width half-maximum,  we
expect an error of 3.8 m\AA , due to the characteristics of our spectra and
their typical $S/N$ ratios  (see Table~\ref{t:reldata}). This value can be
compared with observations.

\begin{figure}
\psfig{figure=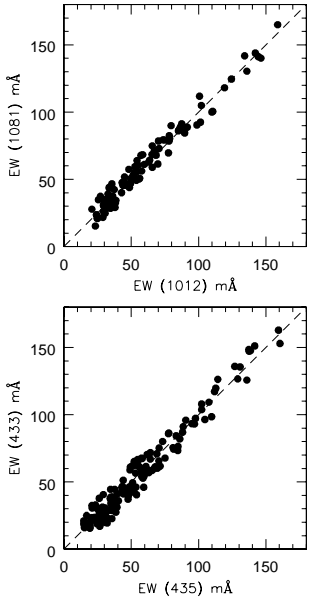,width=8.8cm,clip=}
\caption[]{Comparison between $EW$s measured on the spectra of two dwarfs in 47
Tuc (upper panel; stars 1012 and 1081) and two subgiants (lower panel, stars 433
and 435). The dashed lines indicate equality, and the $rms$ scatters about the
average differences amount to 6.8 m\AA\ in the case of subgiants and
5.5 m\AA\ for TO-stars, in this
example}
\label{f:confew}
\end{figure}

An empirical estimate of internal errors in the equivalent widths can be obtained by
comparing stars of similar physical status and with spectra of similar quality.
We performed this exercise by cross-comparing the sets of $EW$s measured for
subgiants (discarding from the comparison only the 2 stars with lower $S/N$)
and turn-off dwarfs separately.
From 15 comparisons of subgiants, we derived an average $rms$ scatter of 6.4
m\AA\ for the average difference $EW$(star1)-$EW$(star2), based
typically on about 140 lines in common between each pair of stars. In the same
way, the 3 inter-comparisons of the three dwarfs resulted in an average $rms$
scatter of 7.9 m\AA. If we suppose that errors can be attributed in equal
proportions to the stars, we end up with typical errors in $EW$s of 4.5 m\AA\
for subgiants  and 5.6 m\AA\ for the turn-off stars. Figure~\ref{f:confew}
shows two of these comparisons, one for a pair of subgiant stars and one for two
dwarfs.

Comparing the observed typical errors in $EW$s with the expectations of the
Cayrel's formula, we see that another error source is missing, giving residual
uncertainties of 2.5 m\AA\ (quadratic sum) for SGB stars and 4.1 m\AA\ for TO
stars. This can be likely attributed to uncertainties in the location of
the continuum (neglected in the Cayrel formula); with our procedure
(relationship between FWHM and central depth), and considering for simplicity a
triangular shape for the lines, we estimate that errors in the (automatic)
continuum tracing at a level of 1\% (for subgiants) or 2\% (for dwarfs) well
explain the residual discrepancy.

\subsubsection{Evaluation of errors in atmospheric parameters}

Table~\ref{t:sensitivity} shows the sensitivity of the derived abundances  to
 variations in the adopted atmospheric parameters for Fe and other elements;
 this is obtained by
re-iterating the analysis while varying each time only one of the parameters.
This exercise was repeated for a subgiant (star 435) and a dwarf (star 1081).

Note that in some cases the entities of variations listed in
Table~\ref{t:sensitivity} are likely to be overestimates of the actual errors in
the atmospheric parameters. For instance, uncertainties in temperatures were
estimated in Paper I to be $\pm 90$ K for the dwarfs and $\pm 60$ K
for the subgiants. The same values are applicable here.

Errors in the surface gravities can be estimated by taking into account
uncertainties in distance moduli and bolometric corrections (affecting
luminosities: $\pm 0.06$ dex), temperatures ($\pm 0.03$ dex), and masses
($\pm 0.04$ dex). If we sum in quadrature all these contributions, we
find that the adopted gravities have internal errors not larger than 0.08
dex.

Typical errors in the overall metal abundance are less than 0.10 dex,
and have a small effect in the final error budget.

To estimate the proper error bars in the microturbulent velocity values, we
started from the original atmospheric parameters adopted for the subgiant 435 and
the dwarf 1081; then we used the same set of Fe lines to repeat the analysis
changing the $v_t$\ value until the 1$\sigma$ value from the slope of the
abundance/line strength relation was reached.
A simple comparison allows us to
give an estimate of 1$\sigma$ internal errors associated to $v_t$: they are
about 0.7 km/s for subgiants and 0.2 km/s for TO stars.

Col. 7 of Table~\ref{t:sensitivity} allows to estimate the effect of errors in
the $EW$; this was obtained by weighting the average error from a single line
(obtained separately from Fe for the 9 subgiants and the 3 dwarfs) with the
square root of the mean number of lines (listed in Col. 6 of the Table)
measured for each element in subgiants and  dwarf respectively. Finally, the
last column reports the total error bars as the quadratic sum of all the
contributions.

For 47 Tuc we derived from our entire sample of 12 stars an average Fe
abundance of [Fe/H]$=-0.67\pm 0.01 \pm 0.04$ dex, where the first is the
internal error (standard deviation of the mean) and the second represents the
systematic errors, as the weighted average of errors found for TO stars and
subgiants. As stated in Paper I, systematic errors in our case are
mainly due to uncertainties in the temperature scale, which is related to the
scale defined by field stars analyzed in the very same way of our cluster stars.

Note that the Fe solar abundances coming out from the reference analysis made
using the solar model from the Kurucz (1995) grid, with no overshooting, are
slightly different from those used in Paper I, due to the
different treatment of collisional damping (see Gratton et al. 2003b). 
Present adopted
values are $\log n$(Fe)=7.54 for neutral Iron and 7.49 for singly ionized Iron.

We found no difference in the average Fe abundance between subgiant and dwarf
stars in 47 Tuc: [Fe/H]$=-0.67\pm 0.02$ (9 stars) and [Fe/H]$=-0.67\pm 0.01$ (3
stars), respectively. As far as Fe is concerned, 47 Tuc seems to be a very
homogeneous clusters: the $rms$ scatter from star to star is not larger than 0.05
dex, i.e. no more than 12\% for this cluster. This value can be entirely
explained by internal errors, suggesting a very small real star-to-star scatter
in Fe abundances. This scatter well agrees with the upper limit
of $\sim 0.04$ dex estimated by Hesser et al. (1987) from the photometric
intrinsic width of the main sequence in 47 Tuc.

The new average metallicity well agrees (within the quoted uncertainties) with
the value derived in Carretta \& Gratton (1997) from 5 red giants:
[Fe/H]$=-0.70\pm 0.03$. As a further comparison, the revised cluster abundance
based on individual stellar abundances of Fe II, as given recently by Kraft and
Ivans (2003) is [Fe/H]$=-0.63$ dex.
At the face value, the stars in this cluster share the
same amount of Iron in their photospheres, irrespective of their evolutionary
status. A similar result has been obtained for other clusters as well (NGC 6397:
Gratton et al. 2001; M~5: Ramirez and Cohen 2003; M 71: Ramirez and Cohen 2002).

\subsubsection{The ionization and excitation equilibria}

\begin{figure}
\psfig{figure=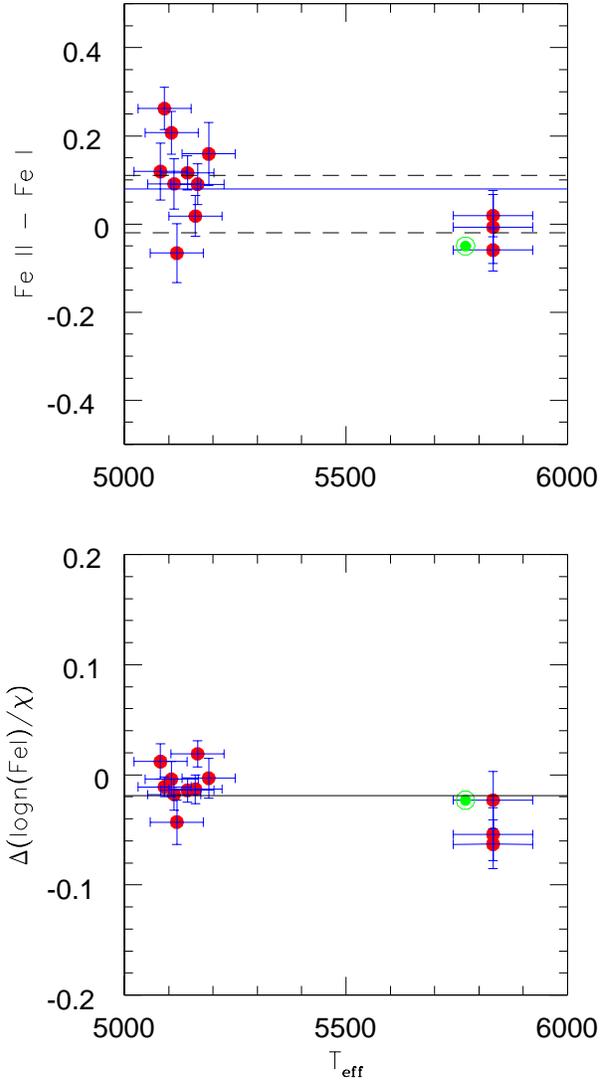,width=8.8cm,clip=}
\caption[]{Upper panel: differences between the abundances of Fe from singly
ionized and neutral lines as a function of effective temperature for stars in
47 Tuc. The solid line is the average difference for the entire sample; the
lower dashed line is the average difference for the 3 dwarfs and the upper
dashed line
represents the average difference for the subgiant stars. The difference for
the Sun, obtained from our reference analysis, is also plotted. Lower panel:
the slopes of the relationship resulting between the Fe I abundances and the
excitation potential as a function of adopted T$_{\rm eff}$'s. The Sun position
in the plot is also marked and the solid line is the average value for our
total sample.}
\label{f:dtheta}
\end{figure}

We can check the uncertainties in Fe abundances using  the
ionization and excitation equilibria as probes, since temperature and gravities were not
derived from our spectra, but independently from colours and theoretical
isochrones (see also the discussion in Carretta et al. 2002).
In Figure~\ref{f:dtheta} we plot the differences between
abundances
obtained from neutral and singly ionized Fe lines and the slopes of the
abundances from neutral lines $\log n$(FeI) with respect to the excitation
potential $\chi$ as a function of effective temperature.

From the upper panel of this Figure we see that on average the lines of singly
ionized Fe give slightly larger (mean difference $+0.08$ dex, $rms$ scatter
$\sigma=0.10$, 12 stars) abundances than neutral lines. The existence of a
trend with temperature is questionable, if we take into consideration the
error bars.
The largest differences are found among the subgiants, whereas the average
difference for the 3 dwarfs is opposite in sign ($\Delta{\rm [Fe/H]}_{\rm II-I} =
-0.02, \sigma=0.03$ dex), and could as well be a reflection of the same
difference in our solar reference values ($-0.05$ dex, see above and
Figure~\ref{f:dtheta}, upper panel).
Do we have to take into account possible departures from the LTE assumption?
The dominating effect being overionization, this could explain the difference
observed among subgiants; however, the good agreement found when comparing
abundances for stars in different evolutionary phases does not support this
hypothesis. On the other hand, errors in atmospheric parameters may affect this
difference.

We started from quoted errors in temperature ($\pm 60$ K and $\pm 90$ K for
subgiants and dwarfs respectively), and used the approximate slope of the
isochrone in the subgiant/dwarf region $\Delta\log g$/$\Delta$T$_{\rm eff}$ (about
0.05 dex/100 K) to estimate the corresponding errors in gravity. We then 
read from Table~\ref{t:sensitivity} the changes in surface gravity and 
overall
metal abundance [A/H] induced by typical errors in temperature, computing the
resulting changes in Fe I and Fe II abundance. The outcome of this exercise is
that the average observed value of $\Delta{\rm [Fe/H]}_{\rm II-I}$ of +0.08 dex (and
more than half of the average difference for subgiants) is reproduced for
temperatures that are lower by 60 (and 90 K) than those presently adopted.
Most of the difference therefore could simply be due to a temperature scale too
low, although well within the error bars.

Another supporting evidence comes from the observed $rms$ scatter (0.10 dex)
observed in the difference $\Delta{\rm [Fe/H]}_{\rm II-I}$. From the last
Column of Table~\ref{t:sensitivity}, the predicted scatters are 0.06 dex and
0.08 dex for subgiant and dwarfs, respectively. Again, a significant fraction
of the observed scatter is explained only by the combination of errors, both in
atmospheric parameters and measured EWs.

Finally, the lower panel of Figure~\ref{f:dtheta} confirms once more that our
adopted temperature scale cannot be grossly wrong, since there is no noticeable
trend of derived abundances as a function of excitation potential. The average
value of the slope $\Delta(\log {\rm Fe})/\chi$ we derive ($-0.019$ dex/eV)
compares well with that derived from the solar analysis
($-0.023$, $\sigma=0.009$ dex/eV) by Carretta et al. (2002), using a very
similar line list.

\begin{table*}
\caption[]{Sensitivities of abundance ratios to errors in the atmospheric
parameters and in the equivalent widths}
\begin{tabular}{lccccrcc}
\hline
Ratio    & $\Delta T_{eff}$ & $\Delta$ $\log g$ & $\Delta$ [A/H] & $\Delta v_t$ &$<N>$& $\Delta$ EW & tot.\\
         & (+100 K)    & (+0.2 dex)      & (+0.2 dex)      & (+0.2 km/s) & &     & (dex)  \\
\\
\hline
& \multicolumn{7}{c}{Star 435 (SGB)} \\
\cline{2-8} \\
$[$O/Na$]$   & $-$0.046& $-$0.005 &$-$0.035 &  +0.003&   &   +0.128& 0.132 \\
$[$O/Fe$]$I  & $-$0.084&   +0.016 &$-$0.071 &  +0.007&  2&   +0.099& 0.119 \\
$[$Na/Fe$]$I & $-$0.038&   +0.021 &$-$0.036 &  +0.004&  3&   +0.081& 0.088 \\
$[$Mg/Fe$]$I & $-$0.046&   +0.007 &$-$0.017 &  +0.005&  3&   +0.081& 0.088 \\
$[$Al/Fe$]$I & $-$0.043&   +0.020 &$-$0.034 &  +0.006&  5&   +0.063& 0.074 \\
$[$Si/Fe$]$I & $-$0.097&   +0.042 &  +0.004 &  +0.004&  3&   +0.081& 0.103 \\
$[$Ca/Fe$]$I & $-$0.010& $-$0.031 &$-$0.007 &  +0.001& 13&   +0.039& 0.043 \\
$[$Sc/Fe$]$II&   +0.035& $-$0.003 &$-$0.002 &$-$0.002&  4&   +0.070& 0.073 \\
$[$Ti/Fe$]$I &   +0.024& $-$0.016 &$-$0.013 &$-$0.001& 11&   +0.042& 0.046 \\
$[$Ti/Fe$]$II&   +0.051& $-$0.042 &  +0.021 &$-$0.001&  5&   +0.063& 0.074 \\
$[$V/Fe$]$I  &   +0.037&   +0.011 &$-$0.030 &  +0.002&  5&   +0.063& 0.069 \\
$[$Cr/Fe$]$I & $-$0.005& $-$0.003 &$-$0.018 &$-$0.001& 10&   +0.044& 0.045 \\
$[$Cr/Fe$]$II& $-$0.005&   +0.002 &$-$0.019 &  +0.000&  4&   +0.070& 0.071 \\
$[$Mn/Fe$]$I & $-$0.004& $-$0.014 &$-$0.006 &$-$0.002&  7&   +0.053& 0.054 \\
$[$Fe/H$]$I  &   +0.097& $-$0.028 &  +0.034 &$-$0.006& 56&   +0.019& 0.068 \\
$[$Fe/H$]$II & $-$0.032&   +0.065 &  +0.066 &$-$0.007&  8&   +0.049& 0.075 \\
$[$Ni/Fe$]$I & $-$0.032&   +0.042 &$-$0.004 &$-$0.001& 10&   +0.044& 0.053 \\
$[$Zn/Fe$]$I & $-$0.090&   +0.032 &  +0.041 &  +0.000&  2&   +0.099& 0.116 \\

\\
\hline
Ratio    & $\Delta T_{eff}$ & $\Delta$ $\log g$ & $\Delta$ [A/H] & $\Delta v_t$ &$<N>$& $\Delta$ EW & tot.\\
         & (+100 K)    & (+0.2 dex)      & (+0.2 dex)      & (+0.2 km/s) & &
	  & (dex)  \\
\\
\hline
& \multicolumn{7}{c}{Star 1081 (TO)}\\
\cline{2-8} \\
$[$O/Na$]$   & $-$0.043 & $-$0.022 &$-$0.028 &  +0.003&   &  +0.219& 0.224\\
$[$O/Fe$]$I  & $-$0.075 &$-$0.032 &$-$0.026 &  +0.033&  3& +0.110& 0.138\\
$[$Na/Fe$]$I & $-$0.032 &$-$0.010 &  +0.002 &  +0.030&  1& +0.190& 0.195\\
$[$Mg/Fe$]$I & $-$0.051 &  +0.003 &$-$0.009 &  +0.034&  2& +0.134& 0.148\\
$[$Si/Fe$]$I & $-$0.053 &  +0.013 &  +0.000 &  +0.028&  1& +0.190& 0.199\\
$[$Ca/Fe$]$I & $-$0.023 &$-$0.002 &$-$0.004 &  +0.015& 11& +0.057& 0.064\\
$[$Sc/Fe$]$II&   +0.018 &  +0.003 &  +0.006 &  +0.026&  3& +0.110& 0.114\\
$[$Ti/Fe$]$I & $-$0.002 &  +0.011 &$-$0.006 &  +0.001&  5& +0.085& 0.085\\
$[$Ti/Fe$]$II&   +0.031 &$-$0.023 &  +0.010 &$-$0.021&  8& +0.067& 0.078\\
$[$V/Fe$]$I  & $-$0.083 &  +0.100 &  +0.023 &  +0.026&  2& +0.134& 0.168\\
$[$Cr/Fe$]$I & $-$0.013 &  +0.016 &$-$0.007 &  +0.002&  8& +0.067& 0.069\\
$[$Cr/Fe$]$II& $-$0.010 &$-$0.001 &$-$0.012 &  +0.021&  3& +0.110& 0.112\\
$[$Mn/Fe$]$I &   +0.000 &$-$0.006 &  +0.006 &  +0.010&  4& +0.095& 0.096\\
$[$Fe/H$]$I  &   +0.090 &$-$0.022 &  +0.009 &$-$0.044& 39& +0.030& 0.105\\
$[$Fe/H$]$II & $-$0.004 &  +0.076 &  +0.030 &$-$0.048&  7& +0.072& 0.110\\
$[$Ni/Fe$]$I & $-$0.022 &  +0.027 &$-$0.005 &  +0.026&  3& +0.110& 0.116\\
$[$Zn/Fe$]$I & $-$0.050 &  +0.042 &  +0.014 &  +0.004&  3& +0.110& 0.123\\
\hline
\end{tabular}
\label{t:sensitivity}
\end{table*}

\section{ANALYSIS AND DERIVED ABUNDANCES}

In the present work we will study only the abundances of the light elements 
O, Na, Al, the group of the $\alpha$-elements and the Fe-groups elements.
Abundances of neutron capture elements in 47 Tuc (as well as in NGC 6752 and
NGC 6397) will be presented in James et al. (2003, in preparation).
The Carbon and Nitrogen abundances, as well as the $^{12}$C/$^{13}$C isotopic
ratios, derived from spectral synthesis of CH and CN molecular bands for stars
in 47 Tuc will be presented and discussed in a forthcoming paper (Carretta et
al. 2003b, in preparation).

\subsection{Lithium}

Also abundances of Lithium in stars of 47 Tuc will be discussed in detail in a
separate, forthcoming paper. Here we point out that abundances of Li and Na
seem to be anticorrelated among 47 Tuc dwarfs. 
The dwarf 
star with the largest equivalent width 
of the Li 6707~\AA\ doublet  (39.1
m\AA) is also the one with  the lowest measured Na abundance
([Na/Fe]$=-$0.34 dex, star 1081, see below). 
The dwarf star with the lowest measured equivalent width of the Li doublet
(14.1 m\AA\ )  is also the dwarf with the highest
Na abundance (star 975, [Na/Fe]=+0.22).
In summary, our preliminary data seem to indicate that in 47 Tuc, turn-off
stars with the largest Na abundance also show 
the largest Li depletion in their
photospheres.

{\tiny
\begin{table*}
\caption[]{Abundances of O, Na and Al in stars of 47 Tuc }
\begin{tabular}{rrrlrclrcr}
\hline
Star   &  n& [O/Fe] & $rms$ & n& [Na/Fe]& $rms$ &
n & [Al/Fe] & $rms$  \\

\hline
\multicolumn{8}{c}{Subgiants}      \\

   435 &2&$<-0.19$& 0.08 &2& +0.31 & 0.12 &4& +0.19&0.10 \\
   456 &1&$<+0.19$&      &3& +0.28 & 0.07 &5& +0.08&0.18 \\
   433 & &        &      &2& +0.24 & 0.05 &5& +0.33&0.10 \\
   478 & &        &      &2& +0.37 & 0.12 &5&$-$0.05&0.19 \\
201600 &2&$<+0.09$& 0.15 &3& +0.30 & 0.02 &6& +0.29&0.12 \\
   429 &2& $-$0.01& 0.12 &2& +0.31 & 0.05 &4& +0.34&0.11 \\
201075 &1&   +0.41&      &3& +0.11 & 0.05 &5& +0.16&0.17 \\
206415 &2&   +0.52& 0.08 &2& +0.10 & 0.12 &5& +0.32&0.05 \\
   482 &3&   +0.61& 0.16 &4& +0.06 & 0.12 &6& +0.42&0.12 \\

\multicolumn{8}{c}{Dwarfs}         \\

1012   &3&   +0.48& 0.01 &1&$-$0.14&      &3&+0.09 & 0.05\\
1081   &3&   +0.57& 0.01 &1&$-$0.34&      &2&$-$0.10&0.07\\
 975   &3&   +0.40& 0.08 &1& +0.22 &      &1&$-$0.19&    \\
\hline
\end{tabular}
\label{t:cno}
\end{table*}
}

\subsection{Oxygen}

Oxygen abundances in these warm stars were derived exclusively from the
permitted near-IR triplet at 7771-75~\AA. In cooler subgiants, only some
of the lines were measured by our automatic procedure, while in warmer dwarfs
all the lines of the triplet were successfully measured. Line measurements were
further checked by eye on the spectra. Only upper limits were derived for three
SGB stars.
Final abundances and upper limits are given in
Table~\ref{t:cno}, corrected for non-LTE effects as described in Gratton et
al. (1999), from statistical equilibrium calculations based on
empirically calibrated collisional H I cross sections. As stated in Carretta
et al. (2000b), these corrections are on average rather small and their
adoption brings the O
abundances derived from the permitted triplet in good agreement with those from
the forbidden [O I] line at 6300~\AA, as far as stars with
T$_{\rm eff} > 4600$ K are considered, which is clearly our case (see also the
discussion in Gratton et al. 2003b).

We carefully looked for the forbidden [O I] line at 6300~\AA\ in  the spectra
of all observed subgiants. Unfortunately, due to the cluster low radial
velocity, in most of stars the [O I] line falls under the telluric emission
line. However, in star 201075 the radial velocity shifts the emission slightly
away, so that the stellar [O I] line could be measured. Using direct
integration, we obtained an EW of 10~m\AA. In order to evaluate the O
abundance, we must take into proper account the contribution of the Ni I line
at 6300.33~\AA\ (see Allende-Prieto et al. 2001 and
Johanson et al 2003). With the adopted atmospheric
parameters and the derived Ni abundance (see below), the Ni line should
contribute for 1~m\AA\ in this star. The resulting EW of 9~m\AA\ for the [O I]
line gives an oxygen abundance of [O/Fe]$=+0.34$ dex, that compares well with
that of +0.41 dex from the IR permitted triplet.

\subsection{Sodium}

47 Tuc is about 0.8 dex more metal-rich than NGC 6752, studied in Paper I. 
Hence, we were able to measure a number of additional Na I lines,
beside those of the strong infrared doublet at 8183-8194~\AA. In fact, lines
of the doublet at 6154-6160~\AA\ were observed in all subgiants and several
measurements of the weak line at 4751~\AA\ were also made.
The doublet at 5682-88~\AA\ is not covered by our spectra.
These lines are
listed as clean in the solar analysis by Holweger (1971).
Corrections for departures from LTE were taken from Gratton et al. (1999),
using collisional cross sections calibrated as in Carretta et al. (2000b).
These corrections are not large (on average,
[Na/Fe]$_{\rm non-LTE}$ - [Na/Fe]$_{\rm LTE}$ = $-0.07\pm 0.01$,
$r.m.s=0.05$ dex for 12 stars), and they improve the agreement between
abundances derived from different indicators. For example, the average
difference in Na abundances obtained from the line at 8183~\AA\ and
those derived as the mean from the doublet lines at 6154-60~\AA\ is only
$-0.07$ dex ($r.m.s=0.06$, 5 stars). Results corrected for the non-LTE effects
are given in Table~\ref{t:cno}.

\subsection{Aluminum}

The strong Al I resonance lines at 3944/61~\AA\ were not measured; they are
heavily saturated and likely affected by large departures from 
LTE (Baum\" uller \& Gehren, 1997). Moreover,
these lines are located on the wings of the strong Ca II H and K lines,
in a region where the  S/N ratios of our spectra is rather low.
Instead, we updated the line list of Paper I (that was limited to the doublet at
6696-98~\AA) including also the doublets at 7835-36~\AA\ and 8772-73~\AA, with
oscillator strengths from Lambert and Warner (1968). Our derived Al abundances
rest on this set of lines.
The non-LTE effects are smaller for high excitation Al lines as the ones
used here (see Ba\" umuller and Gehren 1997). 
Since the amount of corrections for
the subordinate doublets at 6696/6698~\AA, 7835/7836~\AA\ and 8772/8773~\AA\
increases toward $decreasing$ metal abundance, we do not include any correction
for the Al abundances derived for the metal-rich cluster 47 Tuc, even if a
constant offset of about 0.15 dex cannot be excluded (see Table 2 in Baumuller
and Gehren 1997).
Al abundances are summarized in Table~\ref{t:cno}.

\subsection{$\alpha-$elements}

A number of lines for several elements produced by $\alpha$-capture were
measured. Results for Mg, Si, Ca and Ti are given in Table~\ref{t:alpha},
together with the number of lines measured and the $rms$ values obtained for
each species. Abundances from Ti lines in two different stages of ionization
allow us to check for additional evidences of departures from the LTE
assumption. From Table~\ref{t:alpha}, at  face value there is a
small overabundance of singly ionized Ti with respect to the abundances from
neutral lines (on average [Ti/Fe]$_{\rm II}$ - [Ti/Fe]$_{\rm I}$ = 0.12 dex,
with $r.m.s= 0.15$ dex, 12 stars), with a hint of differences decreasing
with increasing temperature among the subgiant stars. However, due to the
rather limited sample size and the small range in T$_{\rm eff}$ covered, we
do not draw any further conclusion.

{\tiny
\begin{table*}
\caption[]{Abundance of $\alpha$-elements in stars of 47 Tuc}
\begin{tabular}{rccccccccccccccc}
\hline
Star   & n& [Mg/Fe]& $rms$ &n& [SiI/Fe]& $rms$ &n& [CaI/Fe]& $rms$ &
n& [TiI/Fe]& $rms$ &n& [TiII/Fe]& $rms$ \\

\hline
\multicolumn{16}{c}{Subgiants}   \\
   435 & 3& +0.31&0.05& 2&+0.33&0.16& 15&+0.18&0.14& 19&+0.37&0.10& 6&+0.36&0.14\\
   456 & 3& +0.32&0.13& 3&+0.30&0.14& 14&+0.24&0.13&  9&+0.31&0.10& 6&+0.47&0.13\\
   433 & 3& +0.49&0.19& 2&+0.21&0.17& 14&+0.22&0.13& 12&+0.17&0.13& 8&+0.25&0.15\\
   478 &  &      &    & 3&+0.10&0.20& 12&+0.11&0.13& 14&+0.14&0.20& 6&+0.58&0.09\\
201600 & 3& +0.33&0.06& 2&+0.23&0.01& 11&+0.24&0.14& 13&+0.27&0.14& 5&+0.50&0.10\\
   429 & 3& +0.39&0.19& 3&+0.42&0.02& 14&+0.09&0.14&  9&+0.17&0.19& 5&+0.28&0.12\\
201075 & 3& +0.42&0.09& 4&+0.38&0.15& 13&+0.20&0.14& 10&+0.37&0.10& 4&+0.49&0.04\\
206415 & 2& +0.34&0.08& 3&+0.45&0.15& 13&+0.27&0.15& 11&+0.24&0.12& 6&+0.31&0.07\\
   482 & 3& +0.64&0.14& 3&+0.28&0.17& 10&+0.23&0.12& 10&+0.32&0.10& 3&+0.15&0.05\\

\multicolumn{16}{c}{Dwarfs}   \\
  1012 & 1& +0.45&    & 1&+0.03&    & 12&+0.13&0.11&  9&+0.21&0.14& 8&+0.35&0.11\\
  1081 & 3& +0.50&0.14& 1&+0.13&    & 12&+0.14&0.11&  6&+0.22&0.14& 8&+0.49&0.18\\
   975 &  &      &   &2&$-$0.10&0.22&  9&+0.21&0.12&  1&+0.06&    &  &     &    \\
\hline
\end{tabular}
\label{t:alpha}
\end{table*}
}

\subsection{Iron-group elements}

Lines of several elements belonging to the Fe-group (Sc, V, Cr, Mn, Ni and Zn)
were measured. Computations of corrections due to the hyperfine structure (HFS)
was performed in the relevant cases (Sc, V and Mn); references are given in
Gratton et al. (2003b).
Another test of the adopted atmospheric parameters and of the LTE assumption
is possible here since we
detected lines of both neutral and singly ionized chromium. The average
difference we found is [Cr/Fe]$_{\rm II}$-[Cr/Fe]$_{\rm I} = -0.07\pm 0.05$
($\sigma=0.16$) dex (11 stars), with no trend with temperature.
Even with large scatter (mainly due to the few lines of ionized chromium
measured), this finding supports our adopted scale for the atmospheric
parameters.

A summary of abundances obtained for the  Fe-group elements is given in
Table~\ref{t:feg1} and Table~\ref{t:feg2}.

{\tiny
\begin{table*}
\caption[]{Abundances of Fe-group elements in stars of 47 Tuc: Sc to Cr}
\begin{tabular}{rcccccccccccc}
\hline
Star   &n&[Sc/Fe]II& $rms$ & n&[V/Fe]I& $rms$ &n&[Cr/Fe]I& $rms$
&n&[Cr/Fe]II& $rms$ \\

\hline
\multicolumn{13}{c}{Subgiants}  \\
   435 &4&$-$0.03&0.20& 4&$-$0.02&0.05& 11& +0.18 &0.12& 4& +0.00  &0.09 \\
   456 &4&  +0.10&0.09& 3&  +0.04&0.04& 10& +0.22 &0.09& 4& +0.27  &0.09 \\
   433 &4&  +0.01&0.11& 5&  +0.02&0.13&  6& +0.14 &0.12& 4&$-$0.05 &0.12 \\
   478 &4&  +0.33&0.13& 5&  +0.09&0.09&  9& +0.03 &0.06& 4& +0.24  &0.15 \\
201600 &4&  +0.32&0.13& 6&$-$0.02&0.16& 11& +0.16 &0.13& 4& +0.19  &0.17 \\
   429 &3&  +0.08&0.14& 7&  +0.11&0.12&  8&$-$0.04&0.11& 4&$-$0.25 &0.17 \\
201075 &2&  +0.04&0.05& 7&  +0.08&0.19& 12& +0.11 &0.09& 3& +0.24  &0.14 \\
206415 &3&  +0.25&0.10& 7&  +0.08&0.13&  9& +0.11 &0.13& 5&$-$0.02 &0.12 \\
   482 &4&  +0.06&0.09& 5&  +0.12&0.17& 10& +0.07 &0.13& 4&$-$0.25 &0.11 \\

\multicolumn{13}{c}{Dwarfs}  \\
  1012 &3&  +0.12&0.12& 2&  +0.10&0.01&  8& +0.10 &0.09& 3&$-$0.04 &0.05 \\
  1081 &3&  +0.12&0.13&  &       &    &  8& +0.08 &0.13& 4& +0.06  &0.07 \\
   975 &2&$-$0.04&0.01&  &       &    &   &       &    & 3& +0.09  &0.11 \\
\hline
\end{tabular}
\label{t:feg1}
\end{table*}
}

{\tiny
\begin{table*}
\caption[]{Abundances of Fe-group elements in stars of 47 Tuc: Mn to Zn}
\begin{tabular}{rccccccccc}
\hline
Star   &n&[Mn/Fe]I & $rms$ & n&[Ni/Fe]I& $rms$ &n&[Zn/Fe]I& $rms$ \\

\hline
\multicolumn{10}{c}{Subgiants}  \\
   435 & 8&$-$0.26&0.07&   7&$-$0.01 &0.11& 2& +0.11 &0.18 \\
   456 & 6&$-$0.18&0.13&   7&  +0.00 &0.12& 2& +0.20 &0.10 \\
   433 & 7&$-$0.40&0.13&  10&  +0.04 &0.15& 2& +0.15 &0.03 \\
   478 & 7&$-$0.24&0.13&   8&  +0.06 &0.15& 2& +0.27 &0.01 \\
201600 & 8&$-$0.28&0.04&   9&  +0.05 &0.14& 2& +0.21 &0.07 \\
   429 & 7&$-$0.39&0.14&  17&  +0.09 &0.18& 2& +0.11 &0.02 \\
201075 & 6&$-$0.16&0.13&  11&  +0.09 &0.12& 3& +0.50 &0.09 \\
206415 & 7&$-$0.37&0.13&  11&  +0.20 &0.14& 2& +0.03 &0.05 \\
   482 & 5&$-$0.30&0.04&  11&  +0.03 &0.14& 2& +0.19 &0.18 \\

\multicolumn{10}{c}{Dwarfs}  \\
  1012 & 6&$-$0.33&0.14&   3&$-$0.25 &0.17& 3& +0.05 &0.12 \\
  1081 & 3&$-$0.51&0.10&   3&$-$0.29 &0.13& 3& +0.06 &0.18 \\
   975 & 4&$-$0.43&0.11&   2&  +0.04 &0.16&  &	     &	   \\
\hline
\end{tabular}
\label{t:feg2}
\end{table*}
}

\section{RESULTS AND DISCUSSION}

\subsection{The Na-O anticorrelation in 47 Tuc}

Mean abundances for individual elements are summarized in
Table~\ref{t:meantab}, where we used only the subgiants, in order to have a more
homogeneous sample of stars all in the same evolutionary phase.
In Col. 3, the average from available
stars (Col. 2) is computed for each element ratio,
and the observed star-to-star scatter is given in Col. 4. The total
uncertainty, as derived from the values of Table~\ref{t:sensitivity}, is
listed in Col. 5.

\begin{table*}
\caption[]{Mean abundance ratios for early subgiant stars in
47 Tuc and in field stars of similar metallicities}
\begin{tabular}{lrccccc}
\hline
Ratio   & N$_{\rm star}$ & Mean & $\sigma_{\rm obs}$ & $\sigma_{\rm tot}$
& Field & $\Delta$ \\
\\
\hline
$[$O/Fe$]$I     &   7&   +0.23 &  0.27 & 0.12 &  +0.53 &$-$0.30\\
$[$Na/Fe$]$I    &   9&   +0.23 &  0.10 & 0.09 &  +0.12 &  +0.11\\
$[$Mg/Fe$]$I    &   8&   +0.40 &  0.10 & 0.09 &  +0.39 &  +0.01\\
$[$Al/Fe$]$I    &   9&   +0.23 &  0.14 & 0.07 &        &\\
$[$Si/Fe$]$I    &   9&   +0.30 &  0.10 & 0.10 &  +0.26 &  +0.04\\
$[$Ca/Fe$]$I    &   9&   +0.20 &  0.06 & 0.04 &  +0.21 &$-$0.01\\
$[$Sc/Fe$]$II   &   9&   +0.13 &  0.13 & 0.07 &  +0.21 &$-$0.08\\
$[$Ti/Fe$]$I    &   9&   +0.26 &  0.08 & 0.05 &  +0.16 &  +0.10\\
$[$Ti/Fe$]$II   &   9&   +0.38 &  0.13 & 0.07 &  +0.16 &  +0.22\\
$[$V/Fe$]$I     &   9&   +0.05 &  0.05 & 0.07 &  +0.07 &$-$0.02\\
$[$Cr/Fe$]$I    &   9&   +0.11 &  0.07 & 0.05 &$-$0.03 &  +0.14\\
$[$Cr/Fe$]$II   &   9&   +0.04 &  0.19 & 0.07 &  +0.06 &$-$0.02\\
$[$Mn/Fe$]$I    &   9& $-$0.29 &  0.08 & 0.05 &$-$0.11 &$-$0.18\\
$[$Fe/H$]$I     &   9& $-$0.67 &  0.05 & 0.07 &        &       \\
$[$Fe/H$]$II    &   9& $-$0.56 &  0.09 & 0.07 &        &       \\
$[$Ni/Fe$]$I    &   9&   +0.06 &  0.06 & 0.05 &$-$0.04 &  +0.10\\
$[$Zn/Fe$]$I    &   9&   +0.20 &  0.13 & 0.12 &  +0.07 &  +0.13\\

\hline
\end{tabular}
\label{t:meantab}
\end{table*}

Note that $\sigma_{\rm tot}$ can be regarded as the predicted total error
expected as a consequence of uncertainties in the adopted atmospheric
parameters, combined with errors in EW measurements. The differences give an
idea of what elements show a scatter that likely exceeds the
predicted observational errors in stars of 47 Tuc.

As one can see from this Table, apart from cases in which the scatter is
large due to the small number of lines measured (like Sc II and Cr II),
two elements clearly stand out: O and Al\footnote{Notice that 
for O the observed
scatter is a lower limit, since 
it was computed including the three subgiants with
upper limit to the O abundance}. 
The scatter observed in Na abundances would be
much larger if the three turn-off stars were considered.

\begin{figure}
\psfig{figure=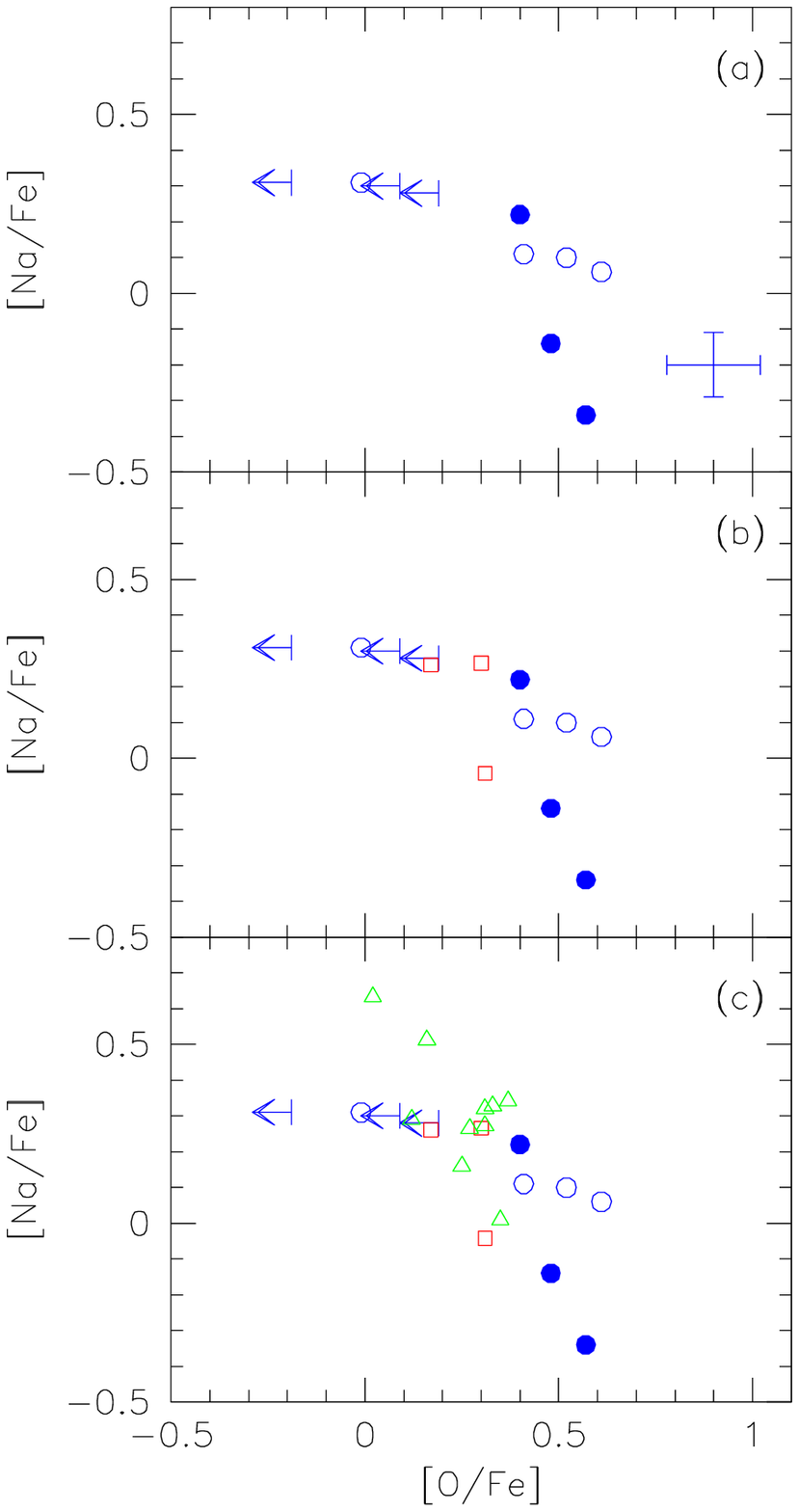,width=8.8cm,clip=}
\caption[]{Run of the [Na/Fe] ratio as a function of [O/Fe]. Panel (a): filled
circles are turn-off stars and open circles are subgiant in 47 Tuc from the
present study. The arrows represent the stars with only upper limits in derived
O abundances. Panel (b): overimposed (open squares) to stars of this study are
3 red giant stars from Carretta (1994). Panel (c): 10 red giants (open
triangles) of the study  of Sneden et al. (1994) in M 71 (NGC 6838), from the
re-analysis by Carretta (1994), are added to the stars of the previous sources.}
\label{f:ona}
\end{figure}

In panel (a) of Figure~\ref{f:ona} we plot the abundance ratios of Oxygen and
Sodium from the present study. The three dwarfs and the subgiants show a well
defined trend, with [O/Fe] abundances anticorrelated with [Na/Fe] ones.
Upper limits in O abundances derived for three other subgiants also contribute
to clearly define the anticorrelation.
{\it This is the first time that the Na-O anticorrelation is detected among
scarcely evolved stars in this cluster.}

In panel (b) of this Figure we added three red giants analyzed by Carretta
(1994). O abundances  for these stars were derived from spectral synthesis of
the forbidden [O I] line at 6300~\AA\ in high resolution spectra acquired with
the CASPEC spectrograph at the ESO 3.6m telescope. Na abundances include the
corrections for departures from LTE, using the same prescription as in Gratton
et al. (1999) and in the present study.

At face value, and despite the small sample size, there seems to be
no significant difference between the pattern shown by giants and less evolved
stars. This impression is confirmed by panel (c) of Figure~\ref{f:ona}, where we
plot as open triangles results from a reanalysis (Carretta 1994) of 10 RGB stars
observed by the Lick-Texas 
group (Sneden et al. 1994) in the globular cluster M 71
(NGC 6838), a ``twin" of 47 Tuc, as far as the overall metal abundance is
concerned. Again, the locus involved in the O-Na anticorrelation seems to be
the same for stars in different evolutionary phases.
We prefer not to use the recent results by Ramirez and Cohen (2002) in order
to plot only Na and O values obtained through homogeneous analyses.
Anyway, their Fig. 11 is in very good agreement with the
present results and let us to conclude that, whatever the phenomenon producing
the Na-O anticorrelation is, it is active also in metal-rich clusters, down to
scarcely evolved stars.

The extension of the Na-O anticorrelation is not as large as that of more
metal-poor cluster, but is undoubtly present.

\begin{figure}
\psfig{figure=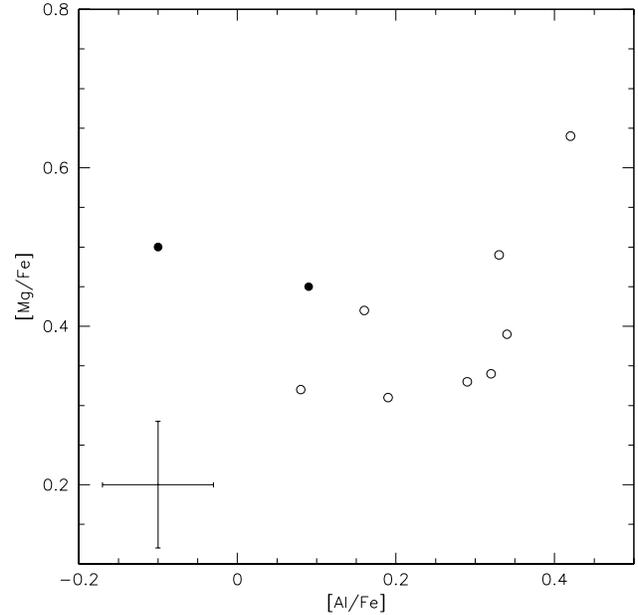,width=8.8cm,clip=}
\caption[]{Run of the [Mg/Fe] ratio as a function of [Al/Fe], for stars in
47 Tuc. Symbols are as in the previous Figure}
\label{f:mgal}
\end{figure}

On the other hand, elements that are produced by p-captures at  higher
temperatures, as Al and Mg, do not show a clear  anticorrelation, as evident in
Figure~\ref{f:mgal}.

The spread in [Mg/Fe] is not astonishing, among the
subgiants, while it is somewhat larger in [Al/Fe]. However, this spread
is only marginally significant (see Table~\ref{t:meantab}), 
slightly exceeding the one expected from uncertainties due to
measurement errors in the EWs and to the effect of adopted atmospheric
parameters\footnote{The possible residual offset due to departure from LTE (see
Sect. 4.4) is constant, and does not affect our conclusion on the
observed spread among subgiants}.
On the other hand, Al is not correlated with Na. 

We note that this is not at odds with the analysis of Al in stars of M 71
(Ramirez and Cohen 2002), where the existing spread could be explained purely
as due to a combination of uncertainties in analysis (see their Tab. 7).  This
could suggest that products of p-captures in the MgAl-cycle are less likely to
occur in metal-rich clusters like 47 Tuc and in M 71. Theoretical models seem
to support this finding: in fact, the efficiency of the Hot Bottom Burning
(HBB) in AGB stars (D'Antona 2003, private communication) is expected to
decreases with increasing metal abundance and the temperature at the base of
the convective envelope is likely not high enough to allow the burning of Mg in
Al in the metal-rich clusters.  Anyway, the limited size of the observed
samples precludes any further conclusion.

In summary, the pattern of light elements observed in unevolved turn-off stars
and in stars at the base of the red giant branch in 47 Tuc seems to be not very
different from that observed in other more metal-poor globular clusters.
The range in Na and O abundances spanned by turn-off stars in 47 Tuc is almost
comparable to that observed by Gratton et al. (2001) in NGC 6752; admittedly,
the number of stars sampled in 47 Tuc is small.

The overall distribution of light elements shows that processes of
proton-capture are at work. In the atmospheres of the studied stars we are
seeing exactly the products of these reactions.
In this case, the line of thought is the same as presented in Gratton et al.
(2001): turn-off stars do not reach the
temperature regime where the ON and NeNa cycles required to produce the Na-O
anticorrelation are active, and moreover these stars have too small convective
envelopes to have efficiently mixed the ashes of these nuclear processing up to
the surface. The same conclusion holds also for subgiants.

The bottom line is that we are seeing products of nuclear
burning and dredge-up in $other$ stars, that are now not observable, but that
returned their elements to the intracluster medium or directly to the surface
of presently observed stars.

\subsection{Self-enrichment in globular clusters?}

\begin{figure}
\psfig{figure=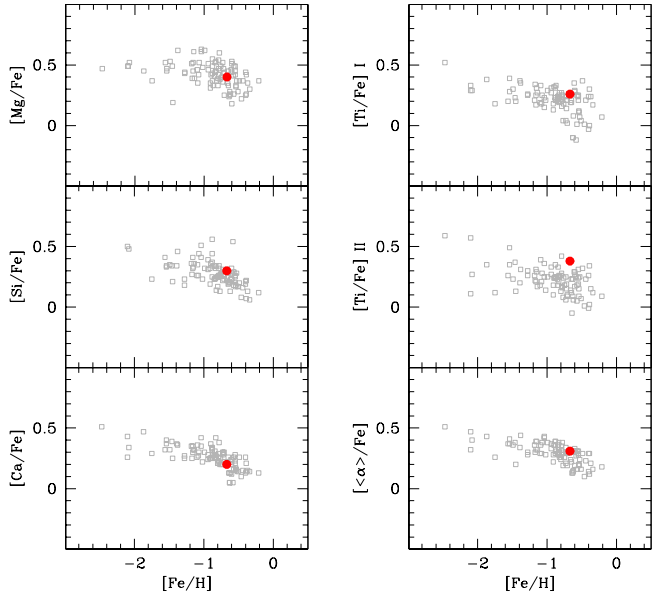,width=8.8cm,clip=}
\caption[]{Run of the overabundances of individual $\alpha-$elements as well as
of the average of Mg, Si, Ca and Ti as a function of [Fe/H]. Open squares are
the field stars belonging to the dissipative component of the Galaxy studied by
Gratton et al. (2003b). The big filled circle represents present data for 47
Tuc. Note the position of the cluster at the upper envelope of the field star
distribution for Ti and the average of $\alpha-$elements. Typical error bars
for the cluster average abundances are smaller than the symbol size.}
\label{f:field}
\end{figure}

Aside from changes in the pattern of light elements,  we might expect that
cluster stars could significantly differ from field stars of the same
metallicity, if the protocluster was able to retain the yields of some
sustained and independent chemical evolution.
Having obtained an extensive set of abundances of $\alpha$-elements in star of
47 Tuc, we may test this hypothesis by looking at the pattern of elements
possibly enriched by the ejecta of core-collapse Supernovae (Cayrel 1986; Brown
et al. 1991, 1995; Parmentier et al. 1999; Parmentier \& Gilmore 2001).
This suggestion is inspired by the fact
that in our own Galaxy the metallicity distribution of GCs is apparently quite
different from that of field stars (e.g., Carney 1993).
Observationally, we might then expect that GCs have a larger [$\alpha$/Fe]
ratio than field stars of similar metal abundance. This signature is likely
subtle, and may have escaped detection in the study by Carney (1996);
it should be more evident among the metal-rich globular clusters,
because observations of field stars indicate that at metallicity
[Fe/H]$\sim -1$\
there is already a significant contribution by type Ia SNe, which has lowered
the [$\alpha$/Fe] ratio from $\sim 0.4$\ to $\sim 0.3$\ (see
Figure~\ref{f:field}).

The elements to be considered here are
those not involved in the O-Na anticorrelation (Mg, Si, Ti, and Ca), for which
very accurate abundances have been obtained. Results for 47 Tuc are compared in
Figure~\ref{f:field} with the overabundances measured in field stars of similar
metallicity in the
dissipative component of our Galaxy (Gratton et al. 2003b). Values are
reported in Col.
6 of Table~\ref{t:meantab}, while in Col. 8 the differences (in the sense
47 Tuc minus field) between the abundances are listed. The analysis is
completely consistent with the present one.

The result for Mg is less significant, since, on average, Mg might be somewhat
reduced due to its participation in the Mg-Al anticorrelation phenomenon,
although there is no evidence that this is the case for 47 Tuc.
Offsets for individual elements are $+$0.04 dex for Si, $-$0.01 dex
for Ca and 0.16 dex for Ti (average of Ti I and Ti II).
On average, 47 Tuc seems to be overabundant in $\alpha$-elements by about $0.06
\pm 0.04$ dex with respect to halo/thick disk stars of similar metallicities.
This result is not significative.

On the other hand, this Figure seems to indicate that some excess of Ti may
indeed be present (a similar indication being obtained for M4 by Ivans et al.
1999), pointing out possibly that GCs produced a significant fraction of their
metals. More extended observations are needed to firmly establish this point.

\subsection{Mn and the enhanced odd-even effect}

A much more puzzling result concerns the abundance of odd Fe-group elements.
Already more than thirty years ago, Helfer et al. (1959) noted that among
metal-poor stars an enhanced odd-even effect for Fe-group elements existed.

\begin{figure}
\psfig{figure=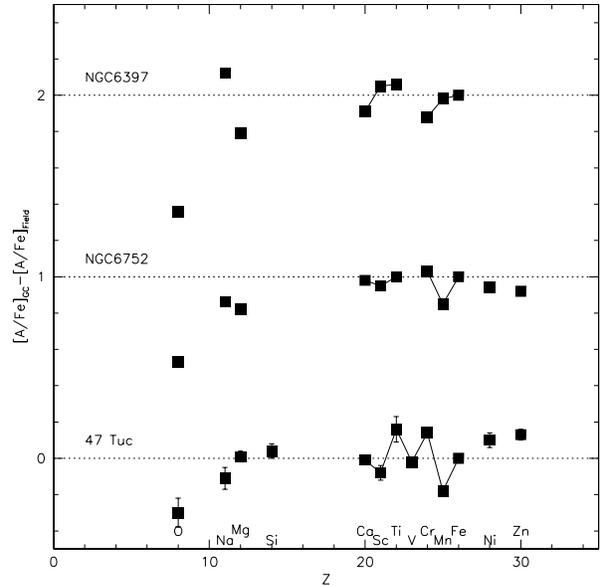,width=8.8cm,clip=}
\caption[]{Differences for several elements between the clusters analyzed in
Paper I and in the present work, and field stars of similar
metallicities (from Gratton et al. 2003b). Differences were arbitrarily shifted
for display purposes. Solid lines connect Fe-group elements having consecutive
atomic number. Notice the enhanced odd-even effect, in particular in 47 Tuc: Sc,
V and Mn are clearly deficient with respect to Ti, Cr and Fe.}
\label{f:oddeven}
\end{figure}

A graphical representation of the odd-even effect in globular clusters is given
in Figure~\ref{f:oddeven}, where the differences
of abundances with respect to field stars of similar metallicities are
displayed for the
three clusters analyzed so far in the Large Program 165.L-0263.
In GCs, the odd-number Fe-group elements show abundances
that are more deficient with respect to the even-number Fe-group elements than
in field stars.
At face value, this odd-even effect seems to increase with increasing
overall metallicity, reaching as much as about 0.2 dex in 47 Tuc. 
This exceeds the observational errors.

Here we concentrate on manganese, 
as the run of this element has been studied since
long time in several stars; moreover, it is the one showing the most marked
underabundance. On the other hand, the major shortcoming is that the
nucleosynthetic site of production of Mn is not yet well known, although 
Mn is
believed to be formed mostly in type Ia SNe (e.g. Nakamura et al. 1999).

Gratton (1989) pointed out that the run of the ratio [Mn/Fe] in metal-poor
stars is mirroring in some way the run of $\alpha$-elements, being almost flat
at very low metallicities, then rising with increasing [Fe/H], starting at
about [Fe/H]$=-1$.
The high, constant [$\alpha$/Fe] ratio in metal-poor stars is commonly
interpreted as due to the delay with whom elements are injected in the
interstellar medium (ISM) as a consequence of different typical lifetimes of the
producers. Massive stars ending their lifes as SN II enrich in very short times
the ISM with $\alpha$-enriched yields; when the longer lived
precursors of SN Ia 
undergo thermonuclear explosion  the major source of
Fe is activated, thus decreasing the ratio [$\alpha$/Fe].
In this framework, Gratton (1989) suggested that the 
observed run of [Mn/Fe] vs
[Fe/H] 
could indicate that the fraction of manganese produced by SN II is lower
than the fraction that SN Ia are allowed to inject in the ISM.

However, the production of elements with odd atomic numbers in the Fe-group
depends on the available neutron excess. Arnett (1971) proposed
that the yields of Mn produced in SN II are then a function of the metallicity,
through the neutron excess. In this view, the rise at higher metallicity would
be due to the occurrence of more metal-rich supernovae.

This approach was recently revived by McWilliam et al. (2003; hereinafter
McW03). They used new observations of giants in the galactic bulge and in the
dwarf spheroidal Sagittarius to constrain the run of [Mn/Fe] at low metal
abundance and at [Fe/H] values typical of the solar neighborhood. Using these
two environments to pinpoint the underabundances of Mn, they claimed that
manganese yields from both SN Ia and SN II are metallicity dependent.

\begin{figure}
\psfig{figure=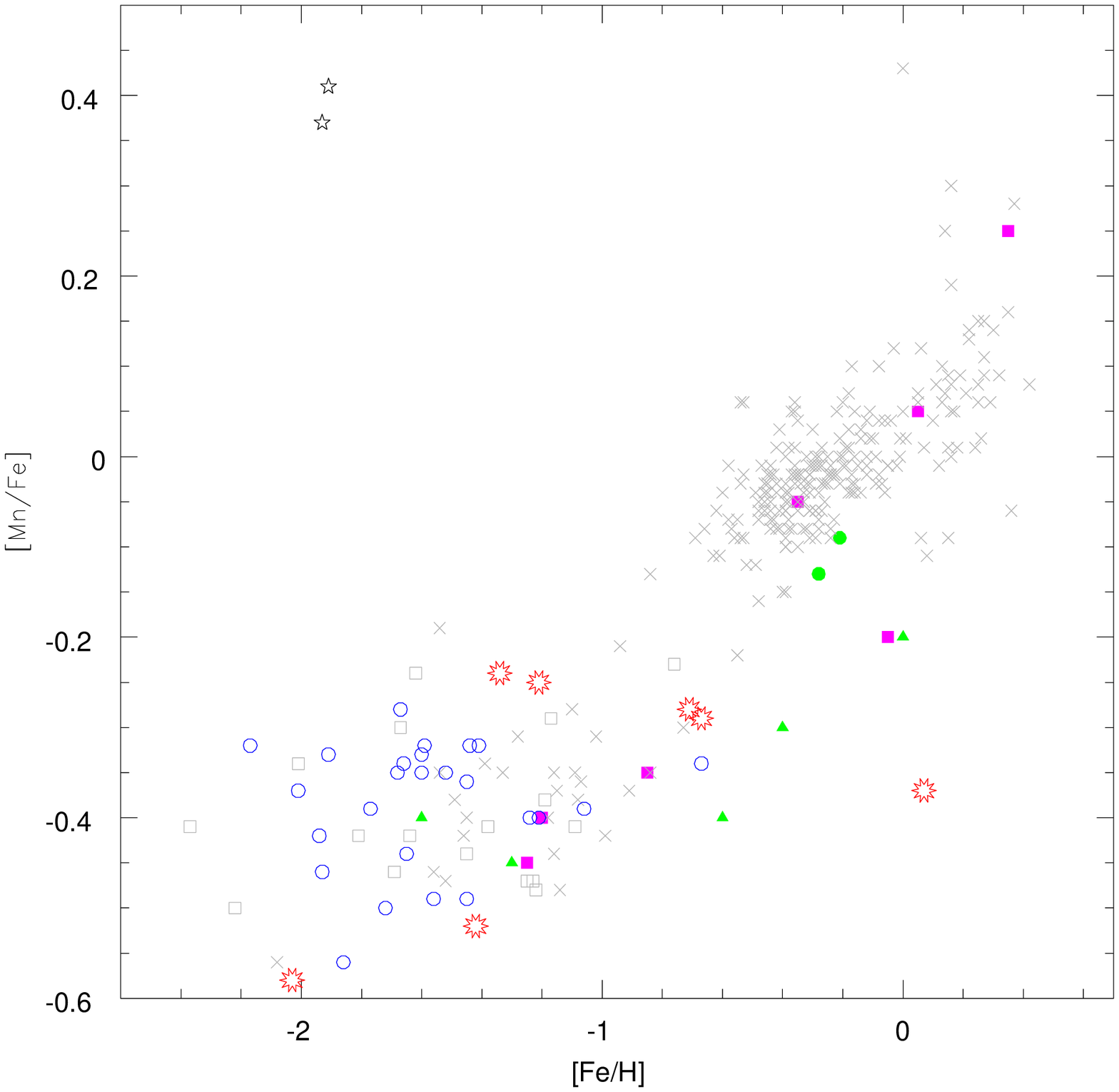,width=8.8cm,clip=}
\caption[]{Run of the [Mn/Fe] ratio as a function of [Fe/H] for
stars in different environments. Light crosses ([Fe/H]$\lsim 0.5$ dex): field
stars of the dissipative component (halo/thick disk) from Gratton
et al. (2003b). Light crosses ($-0.6 \lsim$ [Fe/H] $\lsim 0$): field thin disk
stars from Feltzing and Gustaffson (1998; FG98). Light crosses
([Fe/H]$ \gsim 0$): thin disk stars from Reddy et al. (2000). Open squares:
field stars of the accretion component, from Gratton et al. (2003b). Open
circles: red giants in several dSphs from Shetrone et al. (2001, 2003). Open
five-pointed stars: three low $\alpha$-elements stars from Ivans et al. (2003).
Filled squares: bulge red giants, from McW03. Filled triangles: giants in the
Sgr dwarf galaxy, from McW03. Filled circles: giants in Sgr from Bonifacio et
al. (2000). Open "nova explosions": average abundances for
galactic globular clusters: NGC 6397 and NGC 6752 (Gratton et al. 2001,
[Fe/H]$=-2.03$ and $-1.42$ respectively), Pal 5 (Smith et al. 2002, 
[Fe/H]$=-1.34$), M 5 (Ivans et al. 2001, [Fe/H]$=-1.24$), M 71 (Ramirez and
Cohen 2002, [Fe/H]$=-0.71$), 47 Tuc (this work, [Fe/H]$=-0.67$) and NGC 6528
(Carretta et al. 2001, [Fe/H]=+0.07).}
\label{f:mnvarie}
\end{figure}

However, since their results are based on the still few reliable high
resolution spectra obtainable for these rather distant systems, we gathered Mn
abundances for several kind of environments, to better 
clarify the observational
framework. Results are displayed in Figure~\ref{f:mnvarie}

In order to have a reference, we included field stars from both the accretion
and dissipative components as found by Gratton et al.(2003b), covering the
range of metallicity up to [Fe/H]$\lsim -0.5$. These data were complemented at
higher metallicity with thin disk stars from the study of Reddy et al. (2003),
after an offset of +0.12 dex was added to their [Mn/Fe] ratios to take into
account differences in the adopted solar reference abundances. Also, we plotted
data from Feltzing and Gustaffson (1998), after subtracting 0.04 dex to mantain
the offset of 0.16 dex with Reddy et al. data, as adopted in McW03.

Data points for red giants in Sgr were taken from 
Bonifacio et al. (2000) and  were read from
Figures in the McW03 paper, like those for  the galactic bulge.

Several features in this diagram are worth of discussion.

1) Field stars show the same pattern noted by Gratton (1989): a decreasing run
of [Mn/Fe] for decreasing metallicity, from super-solar values down to about
[Fe/H]$\sim -1.6$, where a sort of plateau, although with large scatter, can be
envisioned around [Mn/Fe]$\sim -0.4$. At  face value, both the quoted
interpretations of SN Ia producing a higher
 fraction of Mn with respect to SN
II and/or more metal-rich SN Ia yielding more Mn than metal-poor SN Ia can be
acceptable to explain this run.

2) Most of the red giants surveyed up to date in dwarf spheroidal galaxies
(dSphs, open circles in Figure~\ref{f:mnvarie}, from Shetrone et al. 2001,
2003) span a range in  metallicity restricted between $-2 <$ [Fe/H] $<-1$, and
the overwhelming majority is scattered around a mean value of [Mn/Fe]$\sim
-0.4$, with no trend whatsoever in increasing the observed [Mn/Fe] ratios over
1 full dex in [Fe/H]. In this respect, we point out that stars in present day
dSphs seem to behave  differently from those in the cannibalized dwarf Sgr (see
below). Our choice of the field reference sample does not affect our
conclusions, that would hold also by using the sample assembled by McW03.

3) Concerning the dSphs, another feature is worth noting:
interestingly enough, the locus occupied by stars in dSphs seems to be
perfectly coincident with that populated by the accretion component selected by
Gratton et al. (2003b), in the [Mn/Fe]-[Fe/H] diagram. This field population is
composed by non-rotating or even counter-rotating halo stars, with distinctive
chemistry with respect to the dissipative component of the halo/thick disk.

Hence, {\it as far as Mn is concerned}, it is tantalizing to conclude that
these stars, likely relics of objects accreted or captured in the past by our
Galaxy, share the same kind of chemical signature of present-day dSphs, i.e.
stellar systems {\it with star formation occurring at a uniformly low rate,
even if at different times} (Tolstoy et al. 2003).

On the other hand, it is well known that the pattern of
$\alpha$-elements differs between stars in dSphs and in the galactic field. The
low (even subsolar) ratios observed in giants stars of present-day dwarfs
suggest that the action of very massive stars was not strong enough to affect
the chemical evolution in these objects (Tolstoy et al. 2003).
The stars of the galactic accretion component (Gratton et al. 2003b), however,
show higher [$\alpha$/Fe] ratios; it is possible that in the
primordial satellites that were likely accreted in the Galaxy, leaving behind
this accreted component, stars with initial masses $> 15-20$ M$_\odot$ were
able to contribute a higher content of $\alpha$-elements.

4) The disrupting dwarf spheroidal Sgr seems to represent a different case. The
five stars analyzed by McW03 and the two by Bonifacio et al. (2000) span a 
much more extended range in metallicity and reach a higher overall metallicity,
maybe reflecting  a higher mass of the host galaxy. Sgr was probably able to
retain a higher amount of gas, reaching up to  solar  values of [Fe/H]. The
most metal poor component of Sgr seems to be intermingled with field stars, and
as well with the most metal-poor component of the bulge population.

Plotting in the same panel both bulge and Sgr stars analyzed by McW03, it 
seems  that  in the high metallicity regime these populations seem to
differentiate from the galactic field population. However considering also the
two stars analyzed by Bonifacio et al. (2000) this distinction seems less
clear; Mn abundances for the larger sample presented by Bonifacio et al. (2003)
will allow to get a clearer view of the situation.  The conclusion of  McW03, 
that also type Ia SNe  produce metallicity-dependent yields, as far as Mn is
concerned, is considerably weakened if the stars of Bonifacio et al. (2000) are
also considered.

5) Concerning the bulge population, the stars analyzed in the paper of McW03
seem to bracket the field population. Using the new data, however, and by
plotting side-by-side bulge and Sgr stars, it is quite
difficult to distinguish the different populations, at the metal poor end. 
In our opinion, in the high
metallicity regime ([Fe/H]$\geq -0.5$), we can only conclude that data encompass
a large range in [Mn/Fe] ratios. This impression is 
strengthened by the average
abundance of the bulge cluster NGC 6528 (Carretta et al. 2001), which is based
on four cluster stars. We therefore caution that 
it may be dangerous to argue about
implications from Mn in metal-rich populations, since the scatter at solar
metallicity is as large as 0.6 dex in [Mn/Fe].

6) The run of globular clusters in the [Mn/Fe] vs [Fe/H] plane seems to be more
or less flat, with some scatter. NGC 6528, as already noted, stands out, since
its [Mn/Fe] ratio is about 0.4 dex below the mean value of stars in the solar
neighborhood. This cluster is also rather peculiar for its kinematics: its high
radial velocity rules out the possibility that NGC 6528 is a disk cluster, but
we cannot exclude that we are just seeing a inner halo cluster presently
passing through the bulge (see discussion in Carretta et al. 2001).
47 Tuc, on the other hand, seems to lie under the mean locus defined by the
field stars at [Fe/H]$\sim -0.7$.

7) Finally, a recent paper by Ivans et al. (2003) shows that the 
situation is
much more complex. They studied in detail three well known $\alpha$-poor
stars (five-pointed stars in Figure~\ref{f:mnvarie}; we dropped in this
Figure star BD+24$^o$ 1676, whose Mn abundance is based on only 1 line). 
While these stars have
the low amount of $\alpha$-elements also shown by stars
in dSphs, 
their Mn contents stand out about 1 dex $above$ the mean level of stars
of similar (very low) metallicity. Moreover, it is likely that these peculiar
stars formed in sub-systems that were distinct from the global galactic
population and subsequently accreted, as hinted by their orbital characteristics
(see Ivans et al. 2003). The bottom line is that the interplay between the
abundances (yields) of $\alpha$-elements and those of Mn is still poorly
understood. As noted by the referee R. Cayrel, the position of these stars
argues that the yield of Mn and Fe are quite different in SN II and SN Ia, in
agreement with theoretical predictions (e.g. Tsujimoto et al. 1995).

8) The situation for Mn is somehow mirrored by that for Ni, an even element:
while in the Galactic field stars Ni tracks iron very closely ([Ni/Fe]$\sim
0.0$) down to extremely low metallicities ([Fe/H]$\sim -4.0$, Cayrel et al.
2003) the GCs Ruprecht 106, Pal 12 (Brown et al. 1997) and Terzan 7 (Sbordone
et al. 2004) as well as the Sgr field stars (Bonifacio et al. 2002) show
[Ni/Fe]$< 0$. Pal 12 and Terzan 7 are in fact  associated to Sgr.  Quite
interestingly in all these objects also [$\alpha$/Fe] is solar or sub-solar. 47
Tuc behaves for Ni like the field stars  and contrasts quite strikingly with 
Terzan 7, which has a similar metallicity. Any theory which seeks to explain
{\it anomalous} Mn abundances should at the same time be able to explain the Ni
abundances.

In this framework, it seems rather difficult to use our derived abundances for
47 Tuc to put strong constraints on the production of Mn.
Summarizing, the commonly adopted
hypothesis are: (i) the contribution from
SN Ia is larger than that from SN II; in this case, our results imply that type
Ia SNe contributed less to the building up of elements in 47 Tuc with respect
to field stars of similar metallicity; and (ii) Mn yields are
metallicity-dependent (via the neutron excess) and the SNe polluting the gas
that originate 47 Tuc were more metal-poor (by about 1 dex) than those
enriching the field stars.

In both cases, however, we might interpret 
these data as hints of a peculiar
chemical history for 47 Tuc. Two tentative scenarios can be envisioned:

(a) the proto-cluster cloud that formed 47 Tuc was able to build up a
significant fraction of its own metals

(b) 47 Tuc might come from a larger object having a chemical evolution quite
distinct from that of our Galaxy. In the scenario of an accretion of a rather
large satellite in the past of our Galaxy (see e.g. Gratton et al. 2000;
Freeman and Bland-Hawthorne 2002), ``inflating" the primordial disk to produce
the present thick disk, 47 Tuc could have formed out of the material of the
infalling satellite during the accretion phase. To support this line of
thought, we note that the age of this cluster is about 2 Gyr less than the
mean age of typical old galactic clusters (Rosenberg et al. 1999;
Gratton et al. 2003a). Moreover,
among the metal-rich ([Fe/H]$>-0.8$) and more massive clusters, 47 Tuc is the
object that 
lies furthest from the galactic plane ($-3$ kpc, from the February 2003
update of Harris 1996).
Besides, there are very few stars of similar metallicity in the dissipative
component whose maximum height above the galactic plane outrank the distance at
which 47 Tuc is presently observed below this plane (Gratton et al. 2003b).

In conclusion, while our data are far from conclusive, we caution that
the observational framework for the Mn abundances is still quite complex and
poorly understood.
Is it
possible that 47 Tuc, one of the most massive globular clusters in the
Galaxy, has been formed in a formerly independent subsystem? While
at present it is not possible to give a clearcut answer with the available
data, we suggest that some hints do exist from both its chemical composition
and orbital features.

\section{SUMMARY}

Using the high resolution UVES spectra we firmly established the
metallicity of 47 Tuc to be [Fe/H]$=-0.67 \pm 0.05$, in substantial agreement
with previous results in the literature. As for the more metal--poor
clusters NGC 6397 and NGC 6752, we find that TO and SGB stars share
the same iron abundance and there is no star-to-star scatter in excess
to what expected from observational errors.

Like in NGC 6752, we have been able to demonstrate the existence of
a Na-O anticorrelation which must be the signature of a previous
generation of stars, since TO and SGB stars do not have high enough
internal temperatures to produce such pattern, no matter
how deep the mixing.

Some scatter has been found also for Li, Mg and Al; no clear
correlations have been found for the last two elements.

For most of the other elements, 47 Tuc seems to behave pretty much 
as field stars of similar metallicity, with two possible exceptions:
Ti seems to be marginally higher than in field stars, Mn seems
to be marginally lower. The errors in our analysis and the scatter
among the field stars are such that none of these anomalies is highly
significant; however, if real, they could point towards the
origin of 47 Tuc as part of an indipendent, possibly larger, system
later captured by the Galaxy. 

\begin{acknowledgements}
{ This research has made use of the SIMBAD data base, operated at CDS,
Strasbourg, France, and was funded by COFIN2001028897 by Ministero Universit\`a
e Ricerca Scientifica, Italy. We wish to thank F. Grundahl for providing
photometry of stars in 47 Tuc. We thank the ESO staff at Paranal (Chile) for
their help during observing runs and Elena Pancino for reading the manuscripts
and making useful comments. }
\end{acknowledgements}


\begin{thebibliography}{}

\bibitem[]{} Alonso, A., Arribas, S. \& Martinez-Roger, C. 1996, A\&A, 313, 873
\bibitem[]{} Allende Prieto, C., Lambert, D.L. \& Asplund, M. 2001, ApJ, 556,
  L63
\bibitem[]{} Arnett, W.D. 1971, ApJ, 166, 153
\bibitem[]{} Baum\"uller, D. \& Gehren, T. 1997, A\&A, 325, 1088
\bibitem[]{} Bell, R.A., Hesser, J.E., Cannon, R.D. 1983, ApJ, 269, 580
\bibitem[]{} Bonifacio P., Hill V., Molaro P., Pasquini L., Di Marcantonio P.,
Santin P.  2000, A\&A , 359, 663 
\bibitem[]{} Bonifacio, P., Sbordone, L., Marconi, G.
 2003, Mem.S.A.It.Suppl., in press 
\bibitem[]{} Bragaglia, A., Carretta, E., Gratton, et al. 2001, AJ, 121, 327
\bibitem[]{} Briley, M.M., Hesser, J.E., Bell, R.A. 1991, ApJ, 373, 482
\bibitem[]{} Briley, M.M., Smith, V.V., Suntzeff, N.B., Lambert, D.L., Bell,
R.A., \& Hesser, J.E. 1996, Nature, 383, 604
\bibitem[]{} Brown, J.H., Burkert, A. \& Truran, J.W. 1991, ApJ, 376, 115
\bibitem[]{} Brown, J.H., Burkert, A. \& Truran, J.W. 1995, ApJ, 440, 666
\bibitem[]{} Brown, 
J.~A., Wallerstein, G., \& Zucker, D. 1997, AJ, 114, 180 
\bibitem[]{} Brown, J.A., Wallerstein, G. 1992, AJ, 104, 1818
\bibitem[]{} Cannon, R.D., Croke, B.F.W., Bell, R.A., Hesser, J.E., \&
 Stathakis, R.A. 1998, MNRAS, 298, 601
\bibitem[]{} Carney, B.W. 1993, in The Globular Clusters-Galaxy Connection, eds.
G.H. Smith and J.P. Brodie, ASP Conf. Ser, 48, 234
\bibitem[]{} Carney, B.W. 1996, PASP, 108, 900
\bibitem[]{} Carretta, E. 1994, Ph.D. Thesis, University of Padova
\bibitem[]{} Carretta, E., \& Gratton, R.G. 1997, A\&AS, 121, 95
\bibitem[]{} Carretta, E., Gratton, R.G., Clementini, G., Fusi Pecci, F. 2000a,
  ApJ, 533, 215
\bibitem[]{} Carretta, E., Gratton, R.G., \& Sneden, C. 2000b, A\&A, 356, 238
\bibitem[]{} Carretta, E., Cohen, J.G., Gratton, R.G. \& Behr, B.B. 2001, 122,
  1469
\bibitem[]{} Carretta, E., Gratton, R.G., Cohen, J.G., Beers, T.C., \&
  Christlieb, N. 2002, AJ, 124, 481
\bibitem[]{} Cayrel, R. 1986, A\&A, 168, 81
\bibitem[]{} Cayrel, R. 1989 in The Impact of Very High S/N Spectroscopy on
Stellar Physics, eds. G. Cayrel de Strobel and M. Spite (Dordrecht:Kluwer), 345
\bibitem[]{} Cayrel, R. et al. 2003, A\&A, in press
\bibitem[]{} Cottrell, P.L., \& Da Costa, G.S. 1981, ApJL, 245, L79
\bibitem[]{} Denisenkov, P.A., Denisenkova, S.N. 1990, Soviet. Astron. Lett., 16, 275
\bibitem[]{} Feltzing, S. \& Gustafsson, B. 1998, A\&AS, 129, 237
\bibitem[]{} Freeman, K. \& Bland-Hawthorn, J. 2002, ARA\&A, 40, 487
\bibitem[]{} Gratton, R.G. 1989, A\&A, 208, 171
\bibitem[]{} Gratton, R.G., Carretta, E., Eriksson, K., \& Gustafsson, B. 1999,
 A\&A 350, 955
\bibitem[]{} Gratton, R.G., Carretta, E., Matteucci, F. \& Sneden, C. 2000,
   A\&A, 358, 671
\bibitem[]{} Gratton, R.G., Sneden, C., Carretta, E., \& Bragaglia, A. 2000,
  A\&A, 354, 169
\bibitem[]{} Gratton, R.G., Bonifacio, P., Bragaglia, A., et al.
2001, A\&A, 369, 87
\bibitem[]{} Gratton, R.G., Bragaglia, A., Carretta, E., Clementini, G.,
  Desidera, S., Grundahl, F., Lucatello, S. 2003a, A\&A, 408, 529
\bibitem[]{} Gratton, R.G., Carretta, E., Claudi, R., Lucatello, S., Barbieri,
  M. 2003b, A\&A, 404, 187
\bibitem[]{} Grundahl, F., Vandenberg, D.A., Stetson, P.B., Anderson, M.I., \&
 Briley, M. 1999, in The Galactic Halo: from Globular
 Clusters to Field Stars (astro-ph/9909447)
\bibitem[]{} Harris, W.E. 1996, AJ, 112, 1487
\bibitem[]{} Helfer, H.L., Wallerstein, G., Greenstein, J.L. 1959, ApJ, 129, 700
\bibitem[]{} Hesser, J.E. 1978, ApJ, 223, L117
\bibitem[]{} Hesser, J.E. \& Bell, R.A. 1980, ApJ, 238, L149
\bibitem[]{} Hesser, J.E., Harris, W.E., Vandenberg, D.A., et al. 1987, PASP,
  99, 739
\bibitem[]{} Holweger, H. 1971, A\&A, 10, 128
\bibitem[]{} Ivans, I.I., Sneden, C., James, C.R., Preston, G.W., Fulbright,
J.P., H\"oflich, P.A., Carney, B.W., Wheeler, J.C. 2003, ApJ, 592, 906
\bibitem[]{} Ivans, I.I., Kraft, R.P., Sneden, C., Smith, G.H., Rich, M.R.,
Shetrone, M. 2001, AJ, 122, 1438
\bibitem[]{} Johansson, S., Litz{\' e}n, 
U., Lundberg, H., \& Zhang, Z.\ 2003, ApJ, 584, L107 
\bibitem[]{} Kraft, R.P. 1994, PASP, 106, 553
\bibitem[]{} Kraft, R.P. \& Ivans, I.I. 2003, PASP, 115, 143
\bibitem[]{} Kurucz, R.L. 1995, CD-ROM 13, Smithsonian Astrophysical
 Observatory, Cambridge
\bibitem[]{} Lambert, D.L. \& Warner, B. 1968, MNRAS, 138, 181
\bibitem[]{} Langer, G.E., Hoffman, R., \& Sneden, C. 1993, PASP, 105, 301
\bibitem[]{} Lucatello, S. \& Gratton, R.G. 2003, A\&A, 406, 691
\bibitem[]{} McWilliam, A., Rich, M.R., Smecker-Hane, T.A. 2003, ApJ, 592, L21
\bibitem[]{} Nakamura, T., Umeda, H., Nomoto, K., Thieleman, F-K., Burrows, A.
   1999, ApJ, 517, 193
\bibitem[]{} Norris, J.E. 1987, ApJ, 313, 65
\bibitem[]{} Norris, J.E., \& Da Costa, G.S. 1995, ApJ, 441, L81
\bibitem[]{} Parmentier, G., Jehin, J., Magain, P., Neuforge, C., Noels, A. \&
Thoul, A.A. 1999, A\&A, 352, 138
\bibitem[]{} Parmentier, G. \& Gilmore, G. 2001, A\&A, 378, 97
\bibitem[]{} Ramirez, S. \& Cohen, J.G. 2002, AJ, 123, 3277 
\bibitem[]{} Ramirez, S. \& Cohen, J.G. 2003, AJ, 125, 224
\bibitem[]{} Reddy, B.E., Tomkin, J., Lambert, D.L. \& Allende Prieto, C. 2003,
   MNRAS, 340, 304
\bibitem[]{} Rosenberg, A., Saviane, I., Piotto, G. \& Aparicio, A. 1999, AJ,
 118, 2306
\bibitem[]{} Sbordone, L., Bonifacio, P., Marconi, G.,
Buonanno, R., 2004, Mem.SAIt , in press 
\bibitem[]{} Shetrone, M.D., C\^ot\'e, P., \& Sargent, W.L.W. 2001, ApJ, 548,
   592
\bibitem[]{} Shetrone, M.D., Venn, K.A., Tolstoy, E., Primas, F., Hill, V. \&
   Kaufer, A. 2003, ApJ, 125, 684
\bibitem[]{} Smith, G.H. 1987, PASP, 99, 67
\bibitem[]{} Smith, G.H., Mateo, M. 1990, ApJ, 353, 533
\bibitem[]{} Smith, G.H., Sneden, C., Kraft, R.P. 2002, AJ, 123, 1502
\bibitem[]{} Sneden, C., Kraft, R.P., Langer, G.E., Prosser, C.F., Shetrone, M.D.
   1994, AJ, 107, 1173
\bibitem[]{} Suntzeff, N.B. 1993, in A.S.P. Conf. Ser., 48, 167
\bibitem[]{} Sweigart, A.V., \& Mengel, J.G. 1979, ApJ, 229, 624
\bibitem[]{} Tassoul, M. \& Tassoul, J.-L. 1984, ApJ, 279, 384
\bibitem[]{} Tolstoy, E., Venn, K.A., Shetrone, M.D., Primas, F., Hill, V.,
 Kaufer, A. \& Szeifert, T. 2003, AJ, 125, 707
\bibitem[]{} Tsujimoto, T., Nomoto, K., Yoshii, Y., Hashimoto, M., Yanagida, S.,
   Thielemann, F.-K. 1995, MNRAS, 277, 945
\end{thebibliography}
\end{document}